\documentstyle[12pt,aaspp4]{article}
\begin{document}
\parindent=1.0cm

\title{The Outer Regions of the Nearby Sc Galaxies NGC 2403 and M33: Evidence 
for an Intermediate Age Population at Large Radii 
\footnote[1]{Based on observations obtained at the
Gemini Observatory, which is operated by the Association of Universities
for Research in Astronomy, Inc., under a co-operative agreement with the
NSF on behalf of the Gemini partnership: the National Science Foundation
(United States), the Particle Physics and Astronomy Research Council
(United Kingdom), the National Research Council of Canada (Canada),
CONICYT (Chile), the Australian Research Council (Australia), CNPq (Brazil),
and CONICET (Argentina).}}

\author{T. J. Davidge}

\affil{Canadian Gemini Office, Herzberg Institute of Astrophysics,
\\National Research Council of Canada, 5071 West Saanich Road,
\\Victoria, B.C. Canada V9E 2E7\\ {\it email: tim.davidge@nrc-cnrc.gc.ca}}

\begin{abstract}

	Deep $g'r'i'z'$ images obtained with the Gemini Multi-Object 
Spectrograph (GMOS) on Gemini North are used to investigate the stellar content 
in the outer regions of the nearby Sc galaxies NGC 2403 and M33. The field 
observed in NGC 2403 covers galactocentric distances between 5 and 11 kpc 
perpendicular to the line of sight (R$_{GC}^{LOS}$), and 7 and 19 kpc along the 
plane of the disk (R$_{GC}^{disk}$). The red giant branch (RGB)-tip occurs 
at $i' = 23.6 \pm 0.1$, and the Cepheid and RGB-tip distance scales 
for NGC 2403 are in good agreement. The number density of bright main 
sequence stars in this field experiences a steep cut-off at R$_{GC}^{disk} \sim 
10$ kpc, which is consistent with the expected truncation radius of the disk 
predicted from studies of edge-on spiral galaxies. While very young 
stars are restricted to R$_{GC}^{disk} <$ 10 kpc, a population of bright 
asymptotic giant branch (AGB) stars is present throughout the 
entire GMOS field, indicating that star formation occured 
outside of the present-day star-forming disk of NGC 2403 during intermediate 
epochs. The AGB stars are not in a tidal stream; in 
fact, the ratio of AGB stars above the RGB-tip to those 
below the RGB-tip does not change with radius, indicating that the bright AGB 
stars are uniformly mixed with the fainter stellar content throughout the 
field. The AGB luminosity function (LF) scales with $r-$band surface brightness 
over a wide range of radii throughout the main body of NGC 2403, 
indicating that the age distribution of stars in the outer regions of the 
present-day star-forming disk is not skewed to younger values than in the 
inner disk. Based on the color of stars on the upper portions of the RGB
it is concluded that metallicity changes across the field, with 
[Fe/H] $= -0.8 \pm 0.1$ (random) $\pm 0.3$ (systematic) at R$_{GC}^{LOS}$ 
= 5 kpc, and [Fe/H] $= -2.2 \pm 0.2$ (random) $\pm 0.8$ 
(systematic) at R$_{GC}^{LOS}$ = 11 kpc. 

	The M33 field samples R$_{GC}^{LOS}$ between 8 and 10 kpc and 
R$_{GC}^{disk}$ between 14 and 17 kpc. Bright AGB stars are detected in this 
field, and the ratio of bright AGB stars to stars on the upper RGB is at 
least as large as that measured in the outer regions of NGC 2403; thus, an 
intermediate-age population occurs well outside of the young star-forming disk 
of M33. The color of stars on the upper RGB suggests that [Fe/H] 
$= -1.0 \pm 0.3$ (random) $\pm 0.3$ (systematic) in this field.

	The globular cluster systems of both NGC 2403 and M33 
contain objects that formed during intermediate epochs, and it is 
suggested that the luminous AGB stars detected in the current study are the 
field counterparts of these clusters. The detection of intermediate-age 
stars in the outer regions of these galaxies is consistent with models in 
which late-type spiral galaxies formed more slowly than earlier-type systems.

\end{abstract}

\keywords{galaxies: individual (NGC 2403, M33) - galaxies: spiral - galaxies: halo - galaxies: stellar content - galaxies: evolution - stars: AGB and post-AGB}

\section{INTRODUCTION}

	The halos and outer disks of spiral galaxies are basic laboratories for 
studying the formation and evolution of these systems. Numerical simulations 
predict that halo formation may be an extended, and perhaps even on-going, 
process that involves the merging of proto-systems having a range of masses and 
chemical enrichment histories (e.g. Bekki \& Chiba 2001). 
The observational evidence that the outer regions of the Milky-Way 
contains stars accreted from companion galaxies is now well established. In 
particular, there are a handful of globular clusters in the Milky-Way that are 
markedly younger than the majority (e.g. Rosenberg et al. 1999; Buonanno et al. 
1998; Sarajedini, Chaboyer, \& Demarque 1997), as well as tidal streams 
containing stars that are distinct from those in the present day halo field 
(e.g. Ibata et al. 2001a; Martinez-Delgado et al. 2001; Helmi \& White 1999; 
Majewski, Munn, \& Hawley 1994); similar structures are also seen in the outer 
regions of M31 (Ibata et al. 2001b; Saito \& Iye 2000). The overall structural 
characteristics of the Galactic halo are also consistent with 
formation driven by hierarchal accretion (e.g. Chiba \& Beers 2001; Bullock, 
Kravtsov, \& Weinberg 2001), although there remain difficulties explaining 
the presence of a thin disk, given that interactions with satellites 
heat, and thereby thicken, the disk (Chiba \& Beers 2001). 

	It is not clear if the age and metallicity distribution function (MDF) 
of halo stars depend on global host galaxy properties such as mass and 
morphology. The stochastic processes inherent to hierarchal evolution 
(e.g. C\^{o}t\'{e} et al. 2000), including differences 
in the gas fraction of accreted satellites (Harris \& Harris 2001), 
are expected to smear any such relations. 
Studies of the brightest resolved stars in the halos 
of Local Group galaxies have revealed intriguing galaxy-to-galaxy differences, 
although the number of systems studied to date is too small to allow secure 
trends to be defined. The halos of M31, the Milky-Way, and M33 have very 
different mean metallicities (e.g. Mould \& Kristian 1986 - hereafter MK86; 
Durrell, Harris, \& Pritchet 1994, 2001), in the sense 
that the M31 halo is the most metal-rich of the three. 
The globular cluster systems of these galaxies also show differences. The 
depths of CN absorption bands in the integrated spectra of M31 and Milky-Way 
clusters are suggestive of different chemical mixtures (Burstein et al. 1984, 
Davidge 1990), while M33 contains a significant number of intermediate age 
globular clusters (Christian \& Schommer 1988 -- hereafter CS88; 
Sarajedini et al. 1998; Chandar et al. 2002) that do not have counterparts in 
either M31 or the Milky-Way. With HST the investigation of resolved stars in 
halos has been expanded to galaxies outside the Local Group. Harris \& 
Harris (2000; 2001) point out that the halo MDF of the elliptical galaxy NGC 
5128, which is more massive than M31, is similar to the MDFs of systems 
that are as diverse as M32 (Grillmair et al. 1996), the LMC (Cole, 
Smecker-Hane, \& Gallagher 2000), and the M31 halo (Durrell et al. 2001), 
which in turn suggests of similarities in their early evolution.

	Disks are thought to form from material that was subjected to tidal 
torques during early epochs (e.g. Zurek, Quinn, \& Salmon 1988). Feedback from 
star formation, coupled with interactions 
between the host galaxy and satellites on plunging orbits, will heat proto-disk 
material and thereby delay disk formation (e.g. Weil, Eke, 
\& Efstathiou 1998). In the case of the Milky-Way, the oldest stars 
in the Galactic disk have ages that are consistently 
lower than those infered for typical globular clusters 
(e.g. Oswalt et al. 1996; Leggett, Ruiz, \& Bergeron 1998; Knox, Hawkins, \& 
Hambly 1999; Liu \& Chaboyer 2000, but see also Binney, Dehnen, \& Bertelli 
2000), which is consistent with a hiatus between the formation of 
the halo and the disk. In fact, the youngest globular 
clusters have ages that are similar to the oldest disk stars, 
suggesting that the formation of the Galactic disk commenced only 
after the conditions required to form globular clusters were suppressed. 

	Disks might be expected to form from the inside out, as 
the low densities of the outer disk make 
these regions more fragile than the inner disk, and hence more easily disrupted 
by disturbances from nearby satellites. The outer disk may also experience 
delayed star formation if infall and the viscous transport of material 
from smaller radii have comparable time scales (e.g. Ferguson \& Clarke 2001). 
The inside-out disk formation model is supported by the presence 
of age gradients in the disks of nearby galaxies, in the sense that mean age 
decreases (becomes younger) towards larger radii (e.g. Bell \& de Jong 2000). 
There is also a tendency for the region near the disk boundary 
in edge-on systems to be bluer than in the rest of the disk, as expected 
if the disk/halo interface is dominated by young stars, although there are some 
exceptions (de Grijs, Kregel, \& Wesson 2001). 

	Deep photometric studies of the resolved stellar contents of nearby 
spiral galaxies offer a direct means of probing the evolution of the outer 
regions of these systems. In this paper deep $g'r'i'z'$ images obtained with 
the Gemini Multi-Object Spectrograph (GMOS) on Gemini North are used to probe 
the bright stellar content in the outer regions of the nearby Sc galaxies M33 
and NGC 2403. The fields observed in these galaxies have similar 
galactocentric radii, R$_{GC}$, thereby permitting a direct comparison of 
stellar contents. 

	NGC 2403 and M33 have roughly the same M$_B$, morphology, inclination, 
gas-phase metal abundances (Garnett et al. 1997), and 
star-forming histories in the inner disk (Davidge \& Courteau 2002); hence, 
they are well matched for a comparative study. 
These galaxies also share a common peculiarity, in that both are 
experiencing vigorous star formation in the inner disk, even though the 
gas densities in these same regions appear to be too low 
to trigger the collapse instabilities required for star formation
(Kennicutt 1989; Martin \& Kennicutt 2001). 
These similarities aside, M33 and NGC 2403 are located in slightly different 
environments. While both galaxies are in a hierarchal system with a larger Sb 
galaxy, NGC 2403 is $\sim 0.9$ Mpc from M81 (Karachentsev et al. 2000), while 
M33 is only $\sim 0.2$ Mpc from M31. Furthermore, the M81 group contains dwarf 
galaxies that do not have known Local Group counterparts (Caldwell et al 1998), 
and this might affect evolution driven by satellite accretion.

	The distance moduli and reddenings adopted for M33 and NGC 2403 
throughout this study are listed in Table 1. The distance modulus for NGC 2403 
in Table 1 is based on Cepheids, and in \S 4 it is demonstrated that 
the Cepheid and RGB-tip distances for NGC 2403 are in good agreement.
As for M33, Cepheids and the RGB-tip give distance moduli 
that differ by 0.3 dex (e.g. Lee et al. 2002), and the RGB-tip distance 
measured by Kim et al. (2002) has been adopted for this study as it is 
likely less susceptible to internal reddening. 

	The observations and data reduction techniques are 
discussed in \S 2, while details of the photometric measurements are 
presented in \S 3. The NGC 2403 field samples a range of environments, and in 
\S 4, 5, and 6 the stellar contents of three distinct regions, that sample the 
outer star-forming disk, the disk/halo interface, and the halo, are discussed. 
The stellar content of the M33 field, which samples a region outside the active 
star-forming disk, is investigated in \S 7. A 
discussion and summary of the results follows in \S 8.

\section{OBSERVATIONS \& REDUCTIONS}

	The data were recorded during Semester 2001B as part of the System 
Verification program for the GMOS (Crampton et al. 2002) on Gemini North. 
$g',r',i',$ and $z'$ images were obtained of fields along the 
minor axes of NGC 2403 and M33; control fields were also observed 
to monitor contamination from foreground stars and background galaxies.
The right ascensions and declinations of the various fields are listed in Table 
2, while the locations of the NGC 2403 and M33 fields on the Palomar Sky Survey 
are shown in Figures 1 and 2. The M33 and NGC 2403 fields have the same 
central R$_{GC}$, although a smaller portion of M33 is 
sampled than NGC 2403 because of the difference in distance. 
In fact, the distance to NGC 2403 is such that the GMOS field samples 
a swath from the star-forming disk to the halo of the galaxy. 

	The data were recorded using a 6 point dither pattern so 
that the gaps between the individual CCDs that make up the 
GMOS detector mosaic could be filled when constructing the final images. Each 
CCD output was binned $2 \times 2$ during read-out, 
so each superpixel covers 0.145 arcsec on a side.
Exposure times were selected using the GMOS Integration Time Calculator (ITC) 
to allow stars on the upper RGB in each galaxy to be detected at the 5 to 
$10 \sigma$ level during median conditions. In general, the 
delivered photometric performance is consistent with the ITC predictions. 
Various details of the observations, including 
exposure time and delivered image quality, 
are listed in Table 2. The exposure times of the NGC 
2403 background field $g'$ observations are much shorter than 
originally intended as the seeing suddenly degraded on the night these data 
were recorded, making it necessary to stop prematurely these observations.

	The data were reduced with tasks in the Gemini IRAF package, using 
bias frames and twilight sky flats that were recorded 
at the beginning and/or end of each night. The data reduction 
sequence consisted of (1) balancing CCD-to-CCD gain 
differences, (2) subtracting the bias pattern and DC bias level, (3) 
dividing by twilight sky flat-field frames, and (4) mosaicking the 
bias-subtracted and flat-fielded outputs from the individual CCD amplifiers 
to create a single image. Interference fringes 
were removed from the $i'$ and $z'$ data by subtracting fringe frames that were
constructed by combining flat-fielded images of various fields recorded through 
these filters. The final images for photometric 
analysis were constructed by spatially aligning the individual images for 
each field, median-combining the results, and then trimming 
to the region of common sky coverage.

	The final $i'$ images of each galaxy and the corresponding control 
field are shown in Figures 3 (NGC 2403) and 4 (M33). The diffuse object near 
the eastern edge of the NGC 2403 control field is the dwarf irregular 
galaxy Kar 50, and the photometric properties of resolved stars 
in this system were the subject of a separate paper (Davidge 2002).

\section{PHOTOMETRIC MEASUREMENTS \& ANALYSIS STRATEGY}

	The photometric measurements were made with the point-spread function 
(PSF) fitting routine ALLSTAR (Stetson \& Harris 1988). PSFs 
were constructed by combining approximately 50 bright stars per image 
with the DAOPHOT (Stetson 1987) PSF task. Because the PSF stars are 
bright and relatively isolated, they were also used 
to determine aperture corrections, which were measured after subtracting 
neighboring sources. The photometric measurements are in the Vega-based 
$g'r'i'z'$ system described by Fukugita et al. (1996), and the calibration was 
set using standard star observations that were obtained during each 
observing run. The estimated uncertainty in the calibration is $\pm 0.05$ mag, 
and this is borne out by the reasonably good agreement with fiducial sequences 
on the $(g'-r', r'-i')$ two-color diagrams (TCDs).

	Artificial star experiments were run to assess 
completeness, estimate photometric uncertainties, and establish a simple means 
of identifying obviously extended objects. This latter step was done by 
rejecting objects having a DAOPHOT `sharp' parameter that differed at the 
$3-\sigma$ or higher level from that predicted for point sources. 
The artificial stars were assigned colors that are typical of real stars in 
the data, and the completeness fractions 
in $g'$ and $i'$ for each field are shown in Figure 5. 

	The NGC 2403 GMOS field contains a mix of populations. 
Young disk stars dominate in the lower right hand corner of 
Figure 1, while older stars dominate in the upper 
left hand corner. Given this gradient in stellar content, the 
NGC 2403 field was divided into three regions, which are indicated in Figure 3. 
The boundary between Regions I and II roughly tracks the drop-off in 
the bright blue main sequence component. Deep imaging studies of edge-on 
spiral galaxies indicate that disks terminate over a $\sim 1$ 
kpc interval (e.g. van der Kruit \& Searle 1981; de Grijs 
et al. 2001), and Region II probes the stellar 
content in a 1 kpc interval beyond the outer edge of Region I. 
Region III occupies the remainder of the GMOS field. The HI disk of 
NGC 2403 extends into Regions II and III (Fraternali et al. 2002).

	The control field was similarly divided so that identical areas are 
covered when making comparisons between the galaxy and control datasets. 
For the remainder of the paper regions in the 
galaxy fields are designated with the subscript `g', while regions in 
the control fields are indicated with the subscript `c'. 

\section{NGC 2403 REGION I$_g$: THE OUTER STAR-FORMING DISK}

\subsection{The LFs, TCDs, and CMDs of NGC 2403 Region I$_g$}

	The $g'$ and $i'$ luminosity functions (LFs) of Region I$_g$, 
corrected for incompleteness and contamination from sources 
in the corresponding portion of the control field, are shown in Figure 6. 
There is a large number of objects belonging to NGC 2403 in Region 
I$_g$, with the onset occuring near $g' = 23.5$ and $i' = 22$. 
The majority of sources with $i' > 23.5$ are evolving on the RGB (\S 4.3).

	The majority of sources in Regions I$_g$ and I$_c$ have stellar 
spectral energy distributions (SEDs), and this is demonstrated in Figures 7 
and 8, where the $(g'-r', r'-i')$ and $(r'-i', i'-z')$ 
TCDs of sources in these regions are compared. The stellar sequence from 
Figure 3 of Schneider et al. (2001, hereafter SDSS) and the log(g) 
= 2.5 and 4.5 solar metallicity models from Tables 9 and 4 of Lenz et al. 
(1998) are also shown in these figures. While the 
majority of sources follow the SDSS sequence on both TCDs when $r'-i' < 0.5$, 
there are stars in Region I$_g$ with $r'-i' > 0.5$ and  
$g'-r' > 1.5$ that fall well off of the SDSS sequence 
on the $(g'-r', r'-i')$ TCD, and stars with similar photometric properties do 
not occur in Region I$_c$. We suggest that the stars with $g'-r' > 1.5$ are red 
giants that are reddened by dust belonging to NGC 2403. That the stars with 
$g'-r' > 1.5$ form a more-or-less continuous sequence 
indicates that the stellar content of Region I$_g$ is not 
subject to uniform levels of internal extinction; rather, there is differential 
extinction ranging up to A$_{g'} \sim 1.0$ mag, which corresponds to 
A$_V \sim 0.9$ and $E(B-V) \sim 0.3$. It is also clear from Figure 
7 that not all giants in Region I$_g$ are affected by large amounts of internal 
extinction. The level of internal extinction predicted from 
the $(g'-r', r'-i')$ TCD is consistent with the range calculated from the 
Balmer decrement in NGC 2403 HII regions (McCall, Rybski, \& Shields 1985; 
Petersen \& Gammelgaard 1996); for comparison, the statistical prescription of 
Tully \& Fouqu\'{e} (1985) predicts a mean internal extinction of A$_B = 
0.34$ mag (A$_V = 0.26$ mag) for NGC 2403 (Pierce \& Tully 1992). 

	The Region I$_g$ and I$_c$ data tend to have $i'-z'$ colors that are 
smaller at a fixed $r'-i'$ than predicted from 
the SDSS sequence. In the case of Region 
I$_g$, the offset with respect to the SDSS data is in the range 0.05 -- 
0.10 mag along the $i'-z'$ axis when $r'-i' < 1$, and so is within the expected 
uncertainty in the photometric calibration. However, when $r'-i' > 1$ there 
is a population of objects that are offset from the SDSS sequence by 0.2 -- 
0.3 mag in $i'-z'$, and hence appear to have non-stellar SEDs. 
This part of the $(r'-i', i'-z')$ TCD is populated by high 
redshift QSOs (Fan 1999), although the density of these objects on the sky is 
too low to produce the number of sources found here (Anderson et al. 2001). 

	The $(r', g'-r')$, $(i', r'-i')$, and $(z', i'-z')$ CMDs of Regions 
I$_g$ and I$_c$ are plotted in Figures 9, 10, and 11. The faint limits of the 
Region I$_g$ and I$_c$ data differ because of differences in image 
quality (e.g. Figure 5). Nevertheless, there is a clear population of stars 
belonging to NGC 2403 in each of the Region I$_g$ CMDs.

\subsection{Young Stars in NGC 2403 Region I$_g$}

	Region I$_g$ contains a population of sources with $g'-r'$ and $r'-i' < 
0$, and the location of these objects on the TCDs indicate that they have 
stellar SEDs. The solid and dashed lines in the left hand panel of Figure 9 
show the main sequence defined by OB stars in the Milky-Way and LMC, translated 
to match the distance to NGC 2403. The main sequence relation 
was obtained by transforming the relation between M$_V$ and effective 
temperature listed by Humphreys \& McElroy (1984) into the SDSS photometric 
system using the relations between (1) $r'$ and $V$ given by Krisciunas, 
Margon, and Szkody (1998) and (2) $g'-r'$ and effective temperature 
predicted by the solar metallicity log(g) = 4.5 Lenz et al. (1998) models. 
The dashed line shows the main sequence reddened only by 
line-of-sight extinction from Schlegel et al. (1998), while the solid line 
shows the main sequence reddened by the foreground and a uniform sheet of 
internal extinction with $E(B-V) = 0.3$, which the distribution of stars on 
the $(g'-r', r'-i')$ TCD suggests is the maximum internal extinction 
for red giants in Region I$_g$ (see above). 

	The amount of reddening in a galaxy is population dependent, in 
the sense that it is higher for young OB stars than for red giants, since stars 
diffuse away from dusty star-forming regions as they age 
(e.g. Zaritsky 1999). The predicted location 
of the main sequence with only foreground reddening lies at the left hand 
edge of the blue stellar content in the Region I$_g$ CMD, indicating that not 
all of the main sequence stars in Region I$_g$ are subject to significant 
internal extinction. The predicted main sequence with internal 
reddening of A$_{g'} = 1$ mag passes through a group 
of stars with $r'$ between 24 and 23, and if these are 
actual main sequence stars then they have intrinsic 
brightnesses comparable to the brightest main sequence stars 
in the Galaxy and the LMC. 

	The brightest blue and red stars in Region I$_g$ are likely 
supergiants. To demonstrate this point, 
the locus defined by type Ia supergiants in 
the Galaxy and Magellanic Clouds, based on the M$_V$ and effective temperatures 
listed in Humphreys \& McElroy (1984) and transformed into the SDSS 
photometric system using the procedure described above for the main 
sequence relation, is also shown on Figure 9 for the two reddening 
situations described above. The Ia supergiant sequence in Figure 9 with 
an internal extinction of A$_{g'} = 1$ mag 
defines the upper envelope of blue stars in Region I$_g$, 
with four stars falling along this sequence. 

	The presence of bright supergiants and main sequence 
stars indicates that Region I$_g$ is a site of recent star formation; 
Region I$_g$ is thus part of the active star-forming disk of NGC 2403.

\subsection{RGB Stars in NGC 2403 Region I$_g$}

	The RGB-tip in moderately metal-poor globular clusters 
occurs near M$_{i'} \sim -4$ and M$_{r'} \sim -2.5$ (Davidge et al. 
2002), which corresponds to $i' \sim 23.5$ and $r' \sim 25$ at the 
distance of NGC 2403 with no internal reddening, and $i' \sim 24.0$ and 
$r' \sim 25.7$ with internal reddening of A$_{g'} = 1$ mag; 
the RGB-tip thus falls near the faint limits of the $(r', g'-r')$ and 
$(z', i'-z')$ CMDs. However, a large population of red stars is seen in the 
$(i', r'-i')$ CMD of Region I$_g$ when $i' > 23.5$.

	The $i'$ LF of sources with $r'-i'$ between 0.2 and 
0.7 in Region I$_g$, which is the color 
range containing the greatest density of RGB stars, is shown in Figure 12. The 
rate of evolution on the RGB is insensitive to metallicity (e.g. VandenBerg 
1992), and so the power-law exponent that characterizes the RGB LF does not 
vary greatly with metallicity. The LF of RGB stars in globular clusters 
follows a power law with $\Delta log(n) / \Delta(i') \sim 0.4$ when [Fe/H] 
$< -1.6$ (Davidge et al. 2002), and a least squares 
fit to the LF entries when $i'$ is between 23.6 and 25.0 indicates that the 
exponent is $0.60 \pm 0.07$, which is higher than that measured in metal-poor 
globular clusters. The fitted relation is compared with the observed LF in 
Figure 12, and there is a tendency for the fitted relation to fall above 
the LF entries when $i' > 23.6$, indicating that the RGB-tip brightness in 
NGC 2403 is $i' = 23.6 \pm 0.1$; to the best of our knowledge, this is the 
first detection of the RGB-tip in NGC 2403. 

	The Cepheid and RGB-tip distance moduli of M33 differ by 0.3 
dex, presumably due to internal extinction (Lee et al. 2002); is this also 
the case for NGC 2403? Using the Cepheid distance modulus for 
NGC 2403 listed in Table 1, and assuming that the majority of RGB-tip stars 
in Region I$_g$ are subject to foreground reddening, then the 
intrinsic brightness of the RGB-tip in Region I$_g$ is M$_{i'}^{RGBT} = 
(23.6 \pm 0.1) - (27.51 \pm 0.24) - (2.086 \times 0.046) = -4.0 \pm 0.3$; 
the last term is the correction for extinction due to forground dust using 
$\frac{A_{i'}}{E(B-V)} = 2.086$ from Table 6 of Schlegel et al. (1998). This 
value of M$_{i'}^{RGBT}$ is consistent with that found in metal-poor globular 
clusters and the dwarf spheroidal galaxy And V by Davidge et al. (2002). 
While the RGB stars in Region I$_g$ are moderately metal-rich (\S 8), the 
RGB-tip brightness in the near-infrared does not change markedly with 
metallicity when [Fe/H] $< -0.7$ (Da Costa \& Armandroff 1990). Thus, 
the Cepheid and RGB-tip distances for NGC 2403 are consistent.

	The mean color and width of the RGB provides information about the 
chemical composition of old and intermediate age populations. In the current 
study, we use $r'-i'$ to probe RGB color, as the data in $r'$ and $i'$ go 
much deeper than the RGB-tip, sample red populations (e.g. Figure 
10), and are less susceptible to line blanketing 
than data recorded at shorter wavelengths (e.g. Bica, Barbuy, \& 
Ortolani 1991). The histogram distribution of $r'-i'$ colors of sources 
in Region I$_g$ with $i'$ between 24 and 25, corrected for sources in the same 
brightness interval in Region I$_c$, is shown in the top panel of Figure 13. 
The analysis is restricted to stars with $i'$ between 24.0 and 25.0 as the $r'$ 
and $i'$ data are 100\% complete at these brightnesses over the range of RGB 
colors. 

	The $r'-i'$ distribution for Region I$_g$ has a 
broad peak, due to the RGB, with a secondary peak when 
$r'-i' < 0$, which is due to bright main sequence stars (\S 4.2). 
The distribution of $r'-i'$ colors about the red peak is symmetric, 
and the mean, based on the data in bins between 0.0 and 1.0, is 
$\overline{r'-i'} = 0.50 \pm 0.01$, where the uncertainty is 
the standard error in the mean. The width of the $r'-i'$ distribution is 
dominated by photometric errors. To demonstrate this point, 
the gaussian error distribution predicted from the artificial star experiments, 
scaled to match the number of stars between $r'-i' = 0.3$ and 0.7, is 
compared with the observed distribution in Figure 13, and there is 
excellent agreement with the observations. Differential reddening of 
size $\Delta$A$_{g'} = 1$ mag does not contribute significantly to the 
width of the observed color distribution. If the standard deviation of the 
internal extinction is one half of the total internal extinction 
(i.e. $\sigma_{A_g'} = \pm 0.5$ mag), then the dispersion along the $r'-i'$ 
axis due to differential reddening is $\sigma_{r'-i'} = \pm 0.1$. The result 
of adding this component in quadrature to that due to photometric errors is to 
broaden the total predicted color distribution by only an additional 
$\pm 0.02$ mag. 

\subsection{AGB Stars in NGC 2403 Region I$_g$}

	AGB stars are seen above the RGB-tip in the Region 
I$_g$ CMDs. Stars near the AGB-tip have low temperatures, and line blanketing 
can suppress the brightnesses of these objects at visible wavelengths, 
causing the AGB to be a near horizontal sequence on CMDs. 
The effects of line blanketing become smaller towards longer wavelengths, and 
the AGB is a vertical sequence on the $(z', i'-z')$ CMD of Region I$_g$. 
Hence, the $z'$ LF of stars above the RGB-tip is an obvious direct probe of AGB 
content. 

	The $z'$ LFs of sources with 
$i' - z'$ between 0 and 1 in Region I$_g$ is shown in Figure 
14. The RGB-tip occurs near $z' = 23.4$, and the onset of the RGB causes the 
Region I$_c$ LF to climb when $z' > 23.4$. The LF is remarkably flat 
between $z'$ = 22.2 and 23.4, and there is an apparent discontinuity at 
$z' = 22$, where the number counts drop. 

	The presence of bright AGB stars above the RGB-tip suggests that Region 
I$_g$ experienced star formation during intermediate epochs. To check this 
result, and also enable comparisons with the AGB contents of other areas of NGC 
2403 and other galaxies, the bolometric LF of AGB stars was constructed 
using brightnesses and colors from the $(i', r'-i')$ CMD of Regions I$_g$ and 
I$_c$. Hudon et al. (1989) used $V, R,$ and $I$ images to investigate the 
AGB content of three fields in NGC 2403 that lie 
roughly along the major axis of the galaxy. 
For consistency, the criteria for defining AGB stars and rejecting red 
supergiants discussed by Hudon et al. (1989) have been adopted for the current 
study. $r'-i'$ colors and $i'$ magnitudes were transformed into $R-I$ and 
$I$ (Kron-Cousins system) using relations from Fukugita et al. (1996) after 
adopting the relation between $V-I$ and $R-I$ for giants from Bessell (1979). 
Bolometric corrections were computed using Equation 2 of Bessell \& Wood (1984).

	The AGB LF of Region I$_g$ 
is shown in Figure 15. The dashed line is the 
composite AGB LF of Fields 2 and 3 from Hudon et al. (1989), which are located
360 and 480 arcsec from the galaxy center. The Hudon et al. (1989) 
LFs in Figure 15 have been (1) scaled to match the number of AGB stars in 
Region I$_g$ with M$_{bol}$ between --4.5 and --6.2, and (2) 
shifted by 0.1 mag along the horizontal axis to match the distance modulus 
adopted here. The dashed-dotted line is the AGB LF of the 
LMC constructed by Reid \& Mould (1984), shifted along the vertical axis to 
match the number of sources in Region I$_g$ with M$_{bol}$ between --4.5 and 
--6.2. Both the Hudon et al. (1989) and Reid \& Mould (1984) LFs match the 
Region I$_g$ LF, suggesting that the star forming histories of Hudon et al. 
(1989) Fields 2 and 3, the LMC, and Region I$_g$ have been similar
during the past few Gyr.

\section{NGC 2403 REGION II$_g$: THE EDGE OF THE YOUNG DISK}

\subsection{The LFs, TCDs, and CMDs of NGC 2403 Region II$_g$}

	The boundary between Regions I and II tracks the drop-off in 
the number of bright blue sources, and so Region II$_g$ does not contain 
objects that are as bright as those in Region I$_g$. This is evident in 
the $g'$ and $i'$ LFs of Region II$_g$, corrected for incompleteness 
and sources in the corresponding portion of the control field, which are 
shown in Figure 6. The onset of stars in Region II$_g$ occurs near $i' = 23$.

	The $(g'-r', r'-i')$ and $(r'-i', i'-z')$ TCDs of Regions II$_g$ and 
II$_b$ are shown in the right hand panels 
of Figures 7 and 8. Whereas the $(g'-r', r'-i')$ TCD of 
Region I$_g$ has an excess population of objects with respect to Region I$_c$ 
over a broad range of colors, the $(g'-r', r'-i')$ TCDs of Regions II$_g$ and 
II$_c$ differ mainly when $g'-r' > 0.6$ and $r'-i' > 0.1$. In fact, the 
$(g'-r', r'-i')$ TCD of Region II$_g$ contains only two objects with $g'-r' 
< 0$ that have stellar SEDs, indicating that the number density of bright 
main sequence stars in Region II$_g$ is lower than in Region I$_g$. An 
absence of sources with $r'-i' < 0$ is also 
evident in the $(r'-i', i'-z')$ TCD of Region II$_g$.

	As was the case in Region I$_g$, the $(g'-r', r'-i')$ TCD of Region 
II$_g$ contains sources with $g'-r' > 1.5$ and $r'-i' > 0.4$ that fall well 
off of the SDSS sequence, whereas Region II$_c$ contains only a modest 
number of objects with similar colors. The presence of objects with $g'-r' 
> 1.5$ suggests that, as in Region I$_g$, some red giants in Region II$_g$ 
are reddened by dust belonging to NGC 2403. 

	The $(r', g'-r')$, $(i', r'-i')$, and $(z', i'-z')$ CMDs of 
Regions II$_g$ and II$_c$ are shown in Figures 9, 10, and 11. 
While some objects that may be young main sequence stars 
are present in the $(r', g'-r')$ CMD of Region II$_g$, 
the brightest of these have $r' > 24$, and hence are 1 mag fainter 
than the brightest main sequence stars in Region I$_g$. While there is a 
star in the $(r', g'-r')$ CMD of Region II$_g$ that has a brightness and 
color consistent with it being a Type Ia supergiant, the number 
density of bright objects in the Region II$_c$ CMD suggests that this 
may be a foreground star with fortuitous photometric properties.

\subsection{RGB Stars in NGC 2403 Region II$_g$}

	In \S 4.3 it was demonstrated that the LF of RGB stars in Region 
I$_g$ follows a power-law with an exponent that is 
larger than that in globular clusters. The $i'$ LF of stars in Regions 
II$_g$ with $r'-i'$ between 0.2 and 0.7, corrected for the number of 
sources in the same color range in Region II$_c$, is shown in Figure 12. 
A least squares fit to sources with $i'$ between 23.6 and 25.0 
in the Region II$_g$ LF indicates that the 
power-law exponent in Region II$_g$ is $\sim 0.27 \pm 0.12$, which 
is smaller than that in Region I$_g$, and is not significantly different 
from that in globular clusters.

	The RGBs of Regions I$_g$ and II$_g$ have different mean colors. 
The $r'-i'$ color distributions of sources with $i'$ between 24 and 25 in 
Region II$_g$, corrected for sources in Region II$_c$, is shown in Figure 13. 
The RGB of Region II$_g$ is skewed towards lower $r'-i'$ values than in Region 
I$_g$, with $\overline{r'-i'} = 0.42 \pm 0.01$. 
Metallicity and extinction are two possible causes of 
the difference between the mean $r'-i'$ colors of upper 
RGB stars in Regions I$_g$ and II$_g$ \footnote[2]{Age is another parameter 
that can affect RGB color. However, in \S 5.3 it is demonstrated that, based on 
the ratio of bright AGB stars to bright RGB stars, the age distributions of 
Regions I$_g$ and II$_g$ during intermediate epochs were likely similar, 
suggesting that the RGBs of Regions I$_g$ and II$_g$ may not have vastly 
different ages.}. While the presence of sources 
with $g'-r' > 1.5$ on the $(g'-r', r'-i')$ TCDs 
of Regions I$_g$ and II$_g$ indicate that both regions contain 
stars that are subject to reddening internal to NGC 2403, Region I$_g$ contains 
sources that are redder than in Region II$_g$, suggesting that the maximum 
reddenings in the two fields may differ, which would cause 
$\overline{r'-i'}$ in Region I$_g$ to be larger than in Region II$_g$. The 
difference in mean giant branch color between Region I$_g$ and II$_g$ is 
$\Delta(r'-i') = 0.08 \pm 0.02$ and, if due entirely to reddening, this 
corresponds to $\Delta A_V = 0.4 \pm 0.1$ mag, which is consistent with the 
difference between the $g'-r'$ colors of the reddest sources in the 
$(g'-r', r'-i')$ TCDs of Regions I$_g$ and II$_g$. However, it is evident 
from the TCDs that not all RGB stars in Regions I$_g$ and II$_g$ are subject 
to this high level of extinction. Moreover, in \S 6.2 it is demonstrated that 
$\overline{r'-i'}$ in Region III$_g$ is lower than in 
Regions I$_g$ and II$_g$ by an amount that is too large to be explained by 
extinction. Extinction is therefore not a convincing explanation for the 
difference in RGB color seen throughout the GMOS field.

\subsection{AGB Stars in NGC 2403 Region II$_g$}

	AGB stars form a spray of objects above the RGB-tip in the Region 
II$_g$ $(i', r'-i')$ and $(z', i'-z')$ CMDs, 
and the presence of these stars indicates that Region II$_g$ experienced star 
formation during intermediate epochs. The overall morphologies of the 
$(z', i'-z')$ CMDs of Regions I$_g$ and II$_g$ are similar: 
as in Region I$_g$, the AGB sequence 
in the $(z', i'-z')$ CMD of Region II$_g$ is almost vertical, 
and peaks near $z' \sim 22$. These similarities suggest that 
Regions I$_g$ and II$_g$ may have similar AGB contents. 

	The $z'$ LF of Region II$_g$ is shown in Figure 14. The $z'$ LFs of 
Regions II$_g$ and I$_g$ are similar, in that both are more-or-less flat in 
the interval between the RGB-tip ($z' \sim 23.3$) and $z' = 22.1$. In fact, 
Regions I$_g$ and II$_g$ have similar numbers of 
bright AGB stars with respect to RGB stars. To demonstrate this point, the 
ratios of the number of sources in Regions I$_g$ and II$_g$ with $z'$ between 
22.1 and 23.3, which are on the AGB, to those with $z'$ between 23.3 and 24.1, 
which are on the upper RGB, R$^{AGB}_{RGB}$, were calculated, 
and the results are given in Table 3. 
R$^{AGB}_{RGB}$ in Region II$_g$ exceeds that in Region I$_g$ at only slightly 
more than the $1-\sigma$ level, indicating that there is not 
a statistically significant difference in the number of AGB stars with repect 
to RGB stars. Given the dearth of young main 
sequence stars in Region II$_g$, the similar values of R$^{AGB}_{RGB}$ 
suggests that intermediate age and very young stars 
in the outer regions of NGC 2403 have different spatial distributions, 
in the sense that the intermediate age population extends out to larger 
R$_{GC}$ than the very young bright main sequence stars. 

	The bolometric LF of AGB stars in Region 
II$_g$, constructed from the $(i', r'-i')$ CMD using 
the procedure discussed in \S 4, is shown in Figure 15. 
As in Region I$_g$, the general shape of the LF is consistent with the Hudon et 
al. (1989) Field 2 $+$ 3 LF, as well as with the LMC LF from Reid \& Mould 
(1984), providing further evidence that the intermediate age 
contents of Regions I$_g$ and II$_g$ are similar. 
The most luminous AGB stars in Region II$_g$ have M$_{bol} = -5.5$, 
which is fainter than in Region I$_g$. However, this may simply be a stochastic 
effect arising from the lower number density of objects in Region II$_g$. 

\section{NGC 2403 REGION III$_g$: THE HALO/OUTER HI DISK}

\subsection{The LFs, TCDs, and CMDs of NGC 2403 Region III$_g$}

	Region III$_g$ samples R$^{disk}_{GC}$ between 15 and 19 kpc, and 
likely contains stars belonging to the halo and the outer HI disk. The 
$g'$ and $i'$ LFs of Region III$_g$, corrected for incompleteness and 
sources in the control field, are shown in Figure 6. The $i'$ LF of 
Region III$_g$ differs from those of Regions I$_g$ and II$_g$ in that the 
onset of the RGB between $i' = 23.5$ and 24.0 is the dominant feature in the 
Region III$_g$ LF, whereas in Regions I$_g$ and II$_g$ the $i'$ LF 
more-or-less continuously climbs towards fainter $i'$. 
There is a modest population of objects in the $i'$ LF 
of Region III$_g$ that extend up to 1.5 mag above the RGB-tip, 
and these are stars evolving on the AGB (\S 6.3).

	Region III$_g$ is dominated by red stars. 
The $(g'-r', r'-i')$ and $(r'-i', i'-z')$ TCDs of Regions III$_g$ and 
III$_c$ are compared in Figures 16 and 17. The number density of sources on the 
$(g'-r', r'-i')$ TCDs of Regions III$_g$ and III$_c$ differ markedly 
when $g'-r' > 0.5$, while there is an absence of objects with stellar SEDs and 
$r'-i' < 0$ and $g'-r' < 0$ in Region III$_g$. 
As in Regions I$_g$ and II$_g$, the $(g'-r', 
r'-i')$ TCD of Region III$_g$ contains stars with 
$g'-r' < 1.5$ and $r'-i' > 0.4$, which are interpreted as giants 
that are reddened by dust internal to NGC 2403. The upper envelope of these 
stars in the Region III$_g$ $(g'-r', r'-i')$ TCD suggests that the maximum 
internal A$_g$ in Region III$_g$ is comparable to that in Region I$_g$.

	The $(r', g'-r'), (i', r'-i')$, and $(z', i'-z')$ CMDs of Regions 
III$_g$ and III$_c$ are shown in Figures 18, 19, and 20. There is a modest 
population of objects in the $(r', g'-r')$ CMD of Region III$_g$ with 
brightnesses and colors that are appropriate for upper main sequence stars. 
However, a comparable number of objects is also seen in the same area of the 
Region III$_c$ $(r', g'-r')$ CMD, indicating that 
the blue objects in Region III$_g$ likely belong to the 
foreground or background. There is also an absence of stars 
in the $(r', g'-r')$ CMD of Region III$_g$ with photometric properties 
suggestive of Type Ia supergiants. Region III$_g$ is thus free of the 
massive young stars associated with the classical star-forming 
disk of NGC 2403. 

\subsection{RGB Stars in NGC 2403 Region III$_g$}

	The $(i', r'-i')$ CMD of Region III$_g$ shows a broad giant branch 
when $i' > 23.5$, with a blue edge near $r'-i' = -0.1$. The 
$i'$ LF of stars with $r'-i'$ between 0.0 and 0.7 in Region III$_g$ is 
shown in Figure 12. A least squares fit to the entries with $i'$ between 23.6 
and 24.8 indicates that the power law exponent of the Region III$_g$ LF is 
$0.60 \pm 0.16$, which is similar to that measured in Region I$_g$, and 
is not significantly different from that measured in globular clusters. 

	The $(r'-i')$ color distributions of stars with $i'$ between 24 and 25 
in the $(i', r'-i')$ CMDs of Region III$_g$, corrected for sources in Region 
III$_c$, is shown in Figure 13. As in Regions I$_g$ and II$_g$, the width of 
the $r'-i'$ distribution in Region III$_g$ is consistent with the observational 
errors predicted from the artificial star experiments. However, the giant 
branch in Region III$_g$ has a mean color $\overline{r'-i'} = 0.33 \pm 0.02$, 
which is significantly smaller than in Regions I$_g$ and II$_g$. Given that 
Regions I$_g$ and III$_g$ appear to have comparable amounts of 
internal differential reddening (\S 6.1), then 
the trend of decreasing $\overline{r'-i'}$ 
with increasing R$_{GC}$ across the GMOS field is likely not due to reddening 
differences, but instead is a consequence of a radial change in stellar content.

\subsection{AGB Stars in NGC 2403 Region III$_g$}

	A population of AGB stars is evident above 
the RGB-tip in the $(i', r'-i')$ and $(z', i'-z')$ CMDs of Region 
III$_g$. The AGB sequence on the $(z', i'-z')$ CMD in Figure 20 is 
morphologically similar to that in Regions I$_g$ and II$_g$. 
The $z'$ LF of stars in Region III$_g$ with 
$i'-z'$ colors between 0 and 1 is shown in Figure 14, and 
the similarity with the LFs of Regions I$_g$ 
and II$_g$ is obvious, both in terms of the peak $z'$ brightness, and the 
shape of the LF. R$^{AGB}_{RGB}$ for Region III$_g$ is listed in Table 3. 
This quantity is similar to what was measured in Regions I$_g$ and II$_g$, 
indicating that the bright AGB stars are uniformly mixed with fainter stars 
between R$_{GC}^{disk} = 7$ and 19 kpc in NGC 2403.

	The bolometric LF of AGB stars in Region 
III$_g$ is shown in Figure 15. As was the case for 
Regions I$_g$ and II$_g$, there is broad agreement 
with the Field $2 + 3$ LF of Hudon et al. (1989), as well as the Reid \& Mould 
(1984) LMC LF. The most luminous AGB stars in Region III$_g$ have M$_{bol} = 
-6$, and so the peak luminosity in Region III$_g$ is similar to that in 
Region I$_g$. The comparisons in Figures 14 and 15 indicate 
that Regions I$_g$, II$_g$, and III$_g$ 
experienced similar star-forming histories during the past few Gyr.

\section{THE HALO AND OUTER HI DISK OF M33}

\subsection{The LFs, TCDs, and CMDs of the Outer Regions of M33}

	With a distance of 0.9 Mpc, M33 is roughly $3 - 4 \times$ closer than 
NGC 2403, and so a single GMOS pointing covers a much smaller 
chunk of this galaxy than NGC 2403. Therefore, the M33 field 
samples a range of projected R$_{GC}$ corresponding only to Regions 
II$_g$ and III$_g$ in NGC 2403. The $g'$ and $i'$ LFs of stars in the M33 
field, corrected for incompleteness and sources in the control 
field, are shown in Figure 6. Contamination from background and foreground 
sources is substantial at all brightnesses, and the number of objects in the 
M33 field exceeds that in the control field only when $g' > 22.5$ and 
$i' > 21$, which corresponds to the upper RGB in M33. 

	Given that M33 and NGC 2403 have similar morphologies, 
inclinations, and integrated brightnesses, it might be 
expected that the number density of red giants in the M33 field should be 
comparable to that in Regions II$_g$ and III$_g$. The numbers of stars in the 
top 1.5 mag in $i'$ of the RGB in the M33 field and the 
three NGC 2403 regions, based on the LFs in Figure 6, are 
summarized in Table 4; the density of stars per square arcsec is listed in 
the last column of this table. Two entries are shown for M33 
-- the first is the observed density of bright giants, while the 
second is the density in this field if it were observed at the distance of NGC 
2403. The latter quantity falls midway between the densities in NGC 2403 
Regions II$_g$ and III$_g$, confirming that the M33 galaxy 
field and the two outermost regions of NGC 2403 have similar stellar densities.

	Sources in the M33 galaxy and control fields have similar SEDs. 
This is demonstrated in Figures 16 and 17, where the $(g'-r', r'-i')$ and 
$(r'-i', i'-z')$ TCDs of sources in these 
fields are compared. There is a modest excess 
population of sources in the lower left hand panels of Figures 16 and 17 with 
respect to the lower right hand panels when $r'-i'$ is between 0 and 1, 
which is the color interval dominated by M33 RGB stars. 
Both fields contain a small number of sources with $g'-r' > 1.5$ 
and $r'-i' > 0.5$. These objects occur in roughly 
equal numbers in the M33 galaxy and control fields. A modest population of 
objects with similar colors are also seen throughout the NGC 
2403 control field, in numbers similar to those in the M33 fields. 
It should be emphasized that the number density of these red objects 
in the control fields is much lower than that of the highly reddened stars that 
were identified in NGC 2403 in \S 4, 5, and 6, and their uniform sky 
distribution suggests that they are background objects. 
While this part of the TCD contains high redshift 
QSOs (e.g. Anderson et al. 2001; Fan 1999), the number density of 
such sources is much lower than that of the red objects found here.

	The $(r', g'-r')$, $(i', r'-i')$ and $(z', i'-z')$ CMDs of the M33 
galaxy and control fields are shown in Figures 18, 19, and 20. The high level 
of contamination from foreground stars and background galaxies is clearly 
evident when comparing the CMDs of the M33 galaxy and control fields, with 
only subtle differences between each pair of CMDs. There is a clear absence of 
bright upper main sequence stars in the $(r', g'-r')$ CMD of M33, and the 
dominant feature in the CMDs is the RGB. The $g'$ observations of M33 sample 
stars that are intrinsically fainter than were observed in NGC 2403, and 
stars on the RGB can be seen in the $(r', g'-r')$ CMD 
with $r' > 22$ and $g'-r'$ between 0.6 and 1.4.

\subsection{RGB Stars in the Outer Regions of M33}

	The RGB of M33 is evident in the $(i', r'-i')$ CMDs in Figure 19 
as a mild increase in stellar density with respect to the control field when 
$i' > 21$. The LFs of stars with $r'-i'$ between 0 and 1 in the M33 field is 
shown in Figure 12. There is only a modest chance of detecting stars in 
each 0.2 mag $i'$ bin when $i' < 21$, and the M33 LF in Figure 12 
contains gaps near the RGB-tip. The modest number of 
bright giants in the M33 field prevents us from measuring the power-law 
exponent of the M33 RGB from these data, so 
a power-law with the exponent fixed at 0.4, which is the 
value measured in globular clusters (Davidge et al. 2002), 
was fit to the LF entries with $i'$ between 21.6 and 22.8, 
and the result is shown in Figure 12. The fitted relation does a 
reasonable job of matching points near the RGB-tip at $i' \sim 20.8$, 
although there appears to be an overall deficiency in the number of giants 
in the upper 0.5 mag of the RGB in $i'$.

	The small number of bright giants in the M33 galaxy field 
complicates efforts to investigate the $r'-i'$ color distribution using the 
procedure employed for NGC 2403; in fact, based on the color distributions in 
Regions II$_g$ and III$_g$, it can be anticipated that only a few stars 
having $i'$ between 21.3 and 22.3 in M33 would be detected per 
0.1 mag bin in $r'-i'$ even near the peak of the color distribution. 
Hence, a wider 0.2 mag binning was used along the color axis 
to investigate the $r'-i'$ color distribution in M33, 
and the results are shown in Figure 13. There 
is an excess population in the M33 galaxy field with respect to the control 
field when $r'-i' = 0.5$, and so it is concluded that 
$\overline{r'-i'} = 0.5 \pm 0.05$ in the M33 galaxy field. 
RGB stars in the M33 halo field are therefore redder than their counterparts in 
NGC 2403 Regions II$_g$ and III$_g$.

\subsection{AGB Stars in the Outer Regions of M33}

	A comparison of the $(z', i'-z')$ CMDs in the lower right and left hand 
panels of Figure 20 indicates that the M33 galaxy field contains bright AGB 
stars. The $z'$ LF of objects with $i'-z'$ between 0 and 1 in the M33 field is 
shown in Figure 14. There are 22 stars in the M33 galaxy field 1 mag above the 
RGB-tip, whereas 10 sources were detected in this same brightness interval 
in the control field. Given the modest number of AGB stars, 
it was decided not to investigate the bolometric LF of these objects. 
Nevertheless, number counts from the $(i', r'-i')$ CMD indicate that 
there is a statistically significant number of stars in the M33 GMOS dataset 
with respect to the control field when M$_{bol} < -4.25$. 

	Some insight into the relative AGB contents of the outer 
regions of M33 and NGC 2403 can be obtained from the 
R$^{AGB}_{RGB}$ statistic, which was computed for the 
three NGC 2403 fields. The number of AGB stars in M33 were counted 
with $z'$ between 19.4 and 20.6, while the number of AGB $+$ RGB stars was 
based on sources having $z'$ between 20.6 and 21.4 to cover the same intrinsic 
brightness intervals used to find R$^{AGB}_{RGB}$ in NGC 2403. 
R$^{AGB}_{RGB}$ for M33 is listed in Table 3. The 
estimated uncertainty in this quantity is substantial due to small number 
statistics; nevertheless, R$^{AGB}_{RGB}$ in the M33 galaxy field 
may be slightly larger than in NGC 2403, 
although the statistical significance is modest. 
While images covering a much larger field of view will be required to determine 
the shape of the AGB LF and the luminosity of the AGB-tip in the outer regions 
of M33, the presence of stars with M$_{bol} < -4$ indicates that, like 
NGC 2403, the outer regions of M33 contain bright AGB stars.

\section{DISCUSSION \& SUMMARY}

	Deep $g'r'i'z'$ images obtained with the GMOS on Gemini North 
have been used to investigate the brightest RGB and AGB stars in the 
inner halo and outer disk of the nearby Sc galaxies M33 and NGC 2403. 
The global properties of these galaxies are 
similar in many respects, and so it might be expected that the stellar 
contents of these systems should also be similar. 
The age distribution of stars in the outer disk of NGC 2403 is discussed 
in \S 8.1, while the metallicities of the M33 and NGC 2403 fields are compared 
in \S 8.2. The nature of the bright AGB stars, which have 
been detected in the outer regions of both galaxies, is discussed in \S 8.3.

\subsection{Comparing the Age Distributions of Stars in the Star-forming Disk of NGC 2403}

	The stellar content in the outermost regions of disks can be used to 
probe the physical processes that drive disk evolution. The age distribution 
of stars is of particular interest, as infall 
and viscosity may feed star-forming material to the outer disk at late 
epochs, and spur star formation in regions that previously 
have not formed stars in large numbers. The age distribution in 
regions of this nature may then be skewed to more recent values than 
in the inner disk (e.g. Ferguson \& Clarke 2001).
Interactions with satellites can also have a major impact 
on the outer disk, creating disruptions 
at large radii (e.g. Laurikainen \& Salo 2001), and possibly causing 
the radial expansion of the disk (Aguerri et al. 2001). Tidal stirring may also 
delay the formation of the outer disk, which has a lower density than the 
inner disk, and hence is more easily disrupted. 

	The AGB is a useful evolutionary phase for probing 
the star-forming history of disks, as it samples stars 
with ages spanning one Gyr or more. In \S 4 it was demonstrated that
the AGB LF of Region I$_g$ is morphologically similar to that measured by Hudon 
et al. (1989) in their Fields 2 and 3, and by Reid \& Mould (1984) in the LMC. 
However, to gain insight into the relative age distributions of these 
regions it is necessary to compare the density of bright AGB stars normalized 
to other stellar objects that trace the mass distribution. One approach is to 
scale the AGB LFs according to total stellar density, which 
can be estimated from surface brightness measurements, such as those 
obtained by Kent (1987a) for NGC 2403 in the $r-$band; a key assumption is 
that the M/L ratio does not change markedly from field-to-field.

	The $r-$band surface brightness measurements for Region I$_g$ 
and each of the Hudon et al. (1989) fields, based on 
the entries in Table II of Kent (1987a), are listed in Table 5. Kent (1987a) 
did not measure the surface brightness of NGC 2403 along the minor axis 
beyond 4.8 arcmin from the galaxy center, 
and so his measurements do not include Region I$_g$, which is 7.5 
arcmin from the center of the galaxy. The surface brightness for 
Region I$_g$ listed in Table 5 was obtained by extrapolating the minor axis 
surface brightness measurements between 2.5 and 4.8 arcmin 
in Table II of Kent (1987a) out to 7.5 arcmin. 

	The AGB LFs of Region I$_g$ and Hudon et al. (1989) Fields 1 and Fields 
2$+$3 are compared in Figure 21, where the Hudon et al. (1989) LFs have been 
scaled to match the surface brightness and areal coverage of Region I$_g$. 
The AGB LFs of Region I$_g$ and Fields $2+3$ agree over the full 
range of luminosities sampled, indicating identical fractional contributions 
from luminous AGB stars to the total stellar content. This in turn implies 
similar age distributions during the time scales probed by AGB stars 
with M$_{bol} < -4.5$, which corresponds to a few Gyr. Region I$_g$ and 
Hudon et al. (1989) Fields $2+3$ sample very different parts of the 
galaxy, and the agreement between the AGB LFs of such spatially diverse 
regions indicates that any stochastic influences on the star-forming history 
in the outer regions of NGC 2403 average out over Gyr timescales.

	The surface brightness measurements indicate that the 
stellar density in Field 1 is $100\times$ that in Region I$_g$. 
The AGB LF of Field 1 agrees within the estimated uncertainties 
with those of Region I$_g$ and Fields $2+3$ when M$_{bol} < -5$, 
indicating that luminous AGB stars account for similar fractions 
of the total $r-$band light in these areas. 
Despite this agreement at the bright end, the Field 1 LF falls below those of 
Region I$_g$ and Fields $2+3$ when M$_{bol} = -4.5$. This may suggest 
that the {\it inner} disk of NGC 2403 experienced a lull in star formation with 
respect to the {\it outer} disk within the past few Gyr; in other words, 
this comparison ostensibly suggests that the age 
distributions of the inner and outer disks during intermediate epochs 
differ. However, it is clear from Figure 6 of Hudon et al. 
(1989) that stars with M$_{bol}$ near --4.5 are at the faint limit 
of their data, raising the possibility that the absence of stars 
with M$_{bol} = -4.5$ in Field 1 may be a consequence of incompleteness. 

	Metallicity gradients are observed in nearby disks (e.g. Zaritsky, 
Kennicutt, \& Huchra 1994; Bell \& de Jong 2000), and the 
gradient in NGC 2403 measured from HII regions is
$\Delta$[O/H]/$\Delta$r = $-0.089 \pm 0.033$ dex kpc$^{-1}$. 
(Zaritsky, Elston, \& Hill 1990). Consequently, at least to the extent 
that [O/H] in HII regions tracks changes in the metallicity of stars formed 
a Gyr in the past, then $\Delta$[O/H] = 0.3 between Region I$_g$ and Hudon 
et al. (1989) Field 1. Changes in metallicity of this size do not greatly 
affect the luminosity of the AGB-tip (e.g. Bertelli et al. 1994), suggesting 
that the difference between the Region I$_g$ and Field 1 LF in Figure 21 at the 
faint end is likely not a consequence of differences in mean metallicity.

\subsection{The Metallicities of the Outer Regions of NGC 2403 and M33}

	The R$^{AGB}_{RGB}$ entries for NGC 2403 Regions I$_g$, II$_g$ and 
III$_g$ in Table 3 are not significantly different, suggesting that these areas 
have similar age distributions, at least during the past few Gyr. If this is 
the case, then the region-to-region differences in the mean $r'-i'$ colors of 
bright RGB stars are most likely due to metallicity. There is as yet no 
empirical or theoretical calibration relating $r'-i'$ to metallicity, and so a 
preliminary calibration is determined here using (1) the cluster giant branch 
sequences of Da Costa \& Armandroff (1990), (2) the relation between $V-I$ and 
$R-I$ for giants from Bessell (1979), and (3) the transformation between 
$r'-i'$ and R-I from Fukugita et al. (1996).

	In keeping with the $r'-i'$ color distributions shown in Figure 13, 
the metallicity calibration is based on the $r'-i'$ color one mag in $i'$ 
below the RGB-tip. The $V-I$ colors of the clusters studied by Da Costa 
\& Armandroff (1990) one mag in $I$ below the RGB-tip are shown in Table 6, 
along with the $R-I$ and $r'-i'$ colors computed with the Bessell 
(1979) and Fukugita et al. (1996) relations. 
Assuming that most of the stars in the NGC 2403 and M33 fields are not 
subject to significant internal extinction, then the 
$r'-i'$ and [Fe/H] entries in Table 6 indicate that the mean metallicities of 
NGC 2403 Regions I$_g$, II$_g$, and III$_g$ are $-0.8 \pm 0.1$ (random) $\pm 
0.3$ (systematic), $-1.4 \pm 0.1$ (random) $\pm 0.4$ (systematic), and 
$-2.2 \pm 0.2$ (random) $\pm 0.8$ (systematic), respectively. 
The random errors reflect the statistical uncertainty 
in $\overline{r'-i'}$, while the systematic errors are due to the $\pm 0.05$ 
mag uncertainty in the photometric calibration. 

	The $r'-i'$ colors computed for the calibrating globular 
clusters in Table 6 rely on color-color relations that are dominated by solar 
neighborhood stars. This is a potential source of systematic uncertainty, 
and so the calibration was checked at both the metal-poor and metal-rich ends. 
The calibration at the metal-poor end was checked directly by comparing the 
observed $r'-i'$ RGB colors of the metal-poor globular cluster NGC 
2419 and the dwarf spheroidal galaxy And V, both of which were discussed by 
Davidge et al. (2002), with the predicted values. The observed 
$r'-i'$ RGB colors of these systems one mag in $i'$ below the RGB-tip 
are listed in the last two rows of Table 6; 
the uncertainty in the photometric calibration of both systems 
is $\pm 0.05$ mag (Davidge et al. 2002). And V and NGC 
2419 have [Fe/H] $= -2.2$, and so should have $r'-i'$ comparable to that of the 
globular cluster NGC 7078, for which $r'-i' = 0.3$ in Table 6. 
Given the $\pm 0.05$ mag uncertainty in the 
photometric calibration of the NGC 2419 and And V data, then the observed RGB 
colors of these systems are consistent with that of NGC 7078.

	The check of the metallicity calibration at the metal-rich end is less 
direct, and relies on HII regions. Disks contain metallicity gradients and so 
only HII regions with R$_{GC}$ comparable to Region I$_g$ are considered. 
There are two HII regions in the Zaritsky et al. (1990) dataset, 
which are \#'s 0 and 29 in their Table I(c), that have R$_{GC}$ close to that 
of NGC 2403 Region I$_g$. Source 0 has log[OIII]/H$\beta$ = 0.91, whereas for 
source 29 this ratio is 0.55. Using the calibration of Edmunds \& Pagel (1984), 
these ratios correspond to $12 +$ log[O/H] $\sim$ 7.9 or 8.1 
for \# 0, and 7.3 or 8.3 for \# 29. \footnote[1]{The relation between 
log[OIII]/H$\beta$ and log(O/H) is multi-valued 
(Edmunds \& Pagel 1984), and so two [O/H] values are 
quoted here.} If $12 +$ [O/H] = 8.9 in the sun (Anders \& Grevesse 1989), then 
Source \# 0 has [O/H] = --0.8 or --1.0, whereas for \# 29 [O/H] = --0.6 or 
--1.6. The upper [O/H] values of these HII regions are in reasonable agreement, 
whereas this is not the case for the lower values, and so the upper 
values are adopted here to estimate [O/H] in Region I$_g$. Taking the 
mean of the upper [O/H] values for these two HII regions gives 
[O/H] $\sim -0.7$ for Region I$_g$, which is 0.3 dex lower than in the LMC 
(Lequeux et al. 1979; Peimbert \& Torres-Peimbert 1974). If the young 
populations in the LMC have [M/H] = --0.3 (Harris \& Zaritsky 2001) and if the 
chemical mixtures in the outer disk of NGC 2403 and the LMC are similar, then 
the young stars in Region I$_g$ would then be 0.3 dex more metal-poor, so that 
[M/H] = --0.6. If the chemical evolution of Region I$_g$ has been similar to 
that of the LMC, which seems reasonable given the similar AGB LFs 
(\S 4), then RGB stars in NGC 2403 will be 0.3 dex more metal-poor 
than the most recently formed stars (Harris \& Zaritsky 2001), so that 
[M/H] = --0.6 $+ -0.3$ = -- 0.9 dex, and this is in excellent agreement with 
what is computed for Region I$_g$ from the $r'-i'$ color of the RGB.

	The metallicity measured for the M33 GMOS field from the $r'-i'$ RGB 
colors is slightly lower than in the corresponding regions of NGC 2403 (i.e. 
Regions II$_g$ and III$_g$), but not significantly so. Based on the calibration 
listed in Table 6 and after accounting for line of sight reddening we find that 
[Fe/H] = $-1.0 \pm 0.3$ (random) $\pm 0.3$ (systematic) for the M33 GMOS 
field. For comparison, the mean metallicity in NGC 2403 Regions II$_g$ 
and III$_g$ is $-1.8 \pm 0.6$; hence, there is not a statistically significant 
difference between the two galaxies.

	The metallicity measured from our M33 galaxy field is larger at the 
$1.4 \sigma$ level than that measured by MK86 in a field with a projected 
distance of 7 kpc from the center of M33. MK86 found that stars in this 
field have a mean metallicity comparable to that of the 
[Fe/H] $\sim -2.2$ cluster M92, and assigned an error of $\pm 
0.8$ dex due to uncertainties in the photometric zeropoint.
A metallicity as low as that measured by MK86 for the halo of M33 
(and that measured here for NGC 2403 Region III$_g$) is 
not consistent with what might be expected from the globular 
cluster system of M33. CS88 studied the ages and metallicities of M33 
clusters, and the six oldest clusters in their compilation, which are those 
with log(t) $\geq 9.6$ in their Table VIII, have [Fe/H] between $-1.2$ and 
$-2.2$, with a mean [Fe/H] $= -1.55 \pm 0.15$. Chandar et al. (2002) 
also find that the majority of halo globular clusters in M33 have [Fe/H] 
between $-1$ and $-2$. Given that globular clusters 
tend to be more metal-poor than their surroundings (e.g. Forbes et al. 1996; 
Da Costa \& Mould 1988), then it can be anticipated that the halo of 
M33 should contain a significant field population with [Fe/H] $\geq -1.6$, 
which is contrary to the MK86 metallicity.

	Cuillandre, Lequeux, \& Loinard (1999) investigated the stellar 
content of a large field in the south east portion of M33, which included 
the area studied by MK86. Based on the RGB sequence in the $(I, V-I)$ CMD, 
Cuillandre et al. (1999) conclude that [Fe/H] $= -1$ with $\pm 0.2$ dex error 
due to uncertainties in the distance to M33. The width of the RGB is 
consistent with a $\pm 0.4 - 0.5$ dex spread in metallicity, 
although the field may be contaminated by 
outer disk stars with a range of ages. Nevertheless, the mean 
metallicity measured by Cuillandre et al. (1999) is consistent with that 
found in the GMOS M33 field and in Region I$_g$ of NGC 2403.

\subsection{Intermediate-age Populations Outside of the Present-Day Star-Forming Disks of NGC 2403 and M33}

	Observations of edge-on spiral galaxies indicate that 
stellar disks terminate on spatial scales of roughly 1 kpc 
(van der Kruit \& Searle 1982; de Grijs, Kregel, \& Wesson 
2001 and references therein), which corresponds to $\sim 1$ arcmin at the 
distance of NGC 2403. The radius at which disk truncatiation occurs is still a 
matter of some debate, and factors such as environment likely 
play a significant role. van der Kruit \& Searle (1982) find that the disk 
cut-off occurs typically at ($4.2 \pm 0.5) \times$ h$_r$, where h$_r$ is the 
radial scale length of the disk, whereas Pohlen, Dettmar, \& L\"{u}tticke 
(2000) find a mean cut-off radius of $(2.9 \pm 0.7) \times$ h$_r$. The disks in 
NGC 2403 and M33 have scale lengths of $2.1 - 2.2$ kpc (Kent 1987a,b), and so 
the stellar disks of these systems might be expected 
to truncate at R$_{GC}^{disk} \sim 6 - 9$ kpc. This 
is roughly consistent with the outer boundary of Region I$_g$, 
where there is a clear drop in the blue main sequence component.

	The presence of bright AGB stars in NGC 2403 Region III$_g$ 
indicates that star formation occured over a much larger range of 
R$_{GC}^{disk}$ in NGC 2403 during intermediate epochs than in the present-day 
star-forming disk. Luminous AGB stars are also present at large R$_{GC}$ 
in the M33 GMOS data (\S 7), as well as in the CMD presented in Figure 4 
of MK86, which contains stars that are redder and brighter than 
the RGB-tip. In fact, there are 29 sources in the MK86 dataset in the 
$I$ interval 1 mag above the RGB-tip, and number counts 
in the M33 GMOS control field predict that only 6 of these are background or 
foreground sources.

	The detection of bright AGB stars at 
large R$_{GC}$ in NGC 2403 and M33 is perhaps not surprising, as 
it has been known for some time that the globular cluster systems of these 
galaxies contain intermediate-age objects at large R$_{GC}$. Battistini et 
al. (1984) used $B-V$ colors to identify three classes of clusters in NGC 2403.
While the majority of blue clusters have small R$_{GC}$, one (C3) is 7 arcmin 
(R$_{GC}^{LOS} \sim 6.4$ kpc) from the galaxy center, while another (F2) is 
15 arcmin (R$_{GC}^{LOS} \sim 14$ kpc) from the center. The M33 globular 
cluster system also contains young and intermediate age clusters 
(CS88, Sarajedini et al. 1998, Chandar et al. 2002,
Ma et al. 2002a,b), and many of these are located at large R$_{GC}$ (e.g. 
Figure 3 of CS88), placing them outside of the main body of the disk. 
The presence of intermediate age globular clusters with halo-like 
kinematics in M33 (e.g. Chandar et al. 2002) indicates that 
the giant molecular clouds that are thought to be the birth sites of 
these systems (e.g. Harris \& Pudritiz 1994; Ashman \& Zepf 2001) were 
present at large R$_{GC}$ in this galaxy or its satellites during intermediate 
epochs.

	M33 and NGC 2403 are not unique among nearby spiral galaxies 
in having intermediate-age populations outside of 
their active star-forming disks, as 
Ferguson \& Johnson (2001) found main sequence stars with a mass of 
3 M$_{\odot}$ in a semi-major axis field 30 kpc from the center of M31. 
However, the outer regions of M31 do not contain the young and intermediate 
age globular clusters seen in M33 and NGC 2403. This could indicate that 
the intermediate-age field populations in the outer regions of M31 
have a smaller fractional size than in M33 and NGC 2403, or that there is a 
difference in origin.

	Did the luminous AGB stars in NGC 2403 and M33 
form in situ, or did they form in a satellite (or satellites) 
that was (were) accreted during recent epochs? An accretion origin 
is attractive because satellite systems at moderately large R$_{GC}$ may 
be able to retain gas up to recent epochs, and hence provide a reservoir of 
material for star formation. However, late satellite accretion 
does not provide an ironclad explanation for the origin of bright 
AGB stars in the outer regions of NGC 2403 and M33. In particular, if the 
halo potentials of these galaxies are near spherical, as predicted by 
simulations (e.g. Bekki \& Chiba 2001), then the rate of 
precession will be modest and tidal streams will not smear 
markedly with time (Ibata et al. 2001). The AGB stars in M33 
and NGC 2403 appear not to be in stream-like structures. In fact,
the uniformity of the quantity R$^{AGB}_{RGB}$, defined in \S 5, with radius in 
NGC 2403, even when including fields that sample the young disk, requires that 
the tidally accreted material would have to be well-mixed within the host 
galaxy.

	The tendency for R$^{AGB}_{RGB}$ not to change with radius in 
NGC 2403 suggests that the bright AGB stars formed in situ. Are the conditions 
in the outer regions of NGC 2403 and M33 at the present epoch such that star 
formation can occur? Star formation is suppressed when the ratio of the local 
gas density to the density required to maintain stability based on the Toomre 
criterion drops below $\sim 0.7$ (Kennicutt 1989; Martin \& Kennicutt 2001). 
Figure 8 of Martin \& Kennicutt (2001) indicates that the gas density in 
NGC 2403 exceeds that required for star formation out to large radii, 
so it presumably should not be difficult to spur 
star formation at large R$_{GC}$ in NGC 2403. 

	Figure 13 of Martin \& Kennicutt (2001) indicates that M33 does not 
contain an extended gas envelope with a density that is high enough to form 
stars according to the standard Kennicutt (1989) threshold. 
Nevertheless, the HI envelope of M33 contains structure (Corbelli, 
Schneider, \& Salpeter 1989), and star formation can occur in regions where the 
mean gas density is lower than the critical value (e.g. Ferguson et al. 
1998), since random perturbations and spiral density waves can cause localized 
gas compression, and thereby trigger star formation (e.g. Kennicutt 1989, 
Ferguson et al. 1998, Martin \& Kennicutt 2001). If star formation is triggered 
by such processes then it might be anticipated that the star-forming 
history in the outermost regions of a galaxy will be less continuous than at 
smaller radii. This being said, the similar AGB contents of 
Regions I$_g$, II$_g$, and III$_g$ and Hudon et 
al. (1989) Fields $2+3$ indicates that star formation 
in the outer most region of NGC 2403 has been coherent over galaxy-wide spatial 
scales when averaged over Gyr time scales.

	Both Regions II$_g$ and III$_g$ sample the outer HI disk of NGC 2403, 
and if the intermediate age stars formed from material from this gas reservoir 
then they might be expected to fall along the plane of the HI disk and have 
disk-like kinematics. Investigations of the number density of bright AGB stars 
in different fields in the outer regions of M33 and NGC 2403, coupled with 
spectroscopic studies to obtain kinematic information, will ultimately provide 
the information needed to chart the spatial distribution of these objects. 
However, there are tantalizing indications that the intermediate 
age globular clusters in late-type spiral galaxies do 
not belong to a disk. For example, Chandar et al. (2002) find 
that many of the intermediate age clusters in M33 do not have 
disk-like kinematics, but appear to belong to a pressure-supported halo. 
Moreover, Kissler-Patig et al. (1999) have found a population of blue clusters 
well off of the disk of the edge-on Sc galaxy NGC 5907.

	Bekki \& Chiba (2001) investigated the formation of halos in 
the context of CMD-based models, and suggested that the amount of halo 
flattening depends on host galaxy morphology, in the sense that late-type 
systems have flatter inner halos. These simulations predict that a flattened 
inner halo in late-type spirals (1) forms well after the more 
spherically distributed outer halo, and (2) contains younger stars, which 
formed during the dissipative merging of gas-rich satellites, than 
the outer halo. The luminous AGB stars that occur 
outside of the actively star-forming disks of M33 and NGC 2403 may be the 
remnants of the final collapse phase of these systems. Such a delayed 
formation model is consistent with the morphological evolution of Sc galaxies 
with redshift, in that these systems are observed to become increasingly 
chaotic at intermediate redshifts (van den Bergh 2002). 

\subsection{Future Work}

	There is an obvious need to investigate additional fields sampling a 
larger range of R$_{GC}$ in M33 and NGC 2403, as studies of the spatial 
distribution of AGB stars in the outer regions of these galaxies will provide 
clues about the origin of intermediate age populations 
at large R$_{GC}$. Photometric and spectroscopic surveys 
of the M33 and NGC 2403 globular cluster systems will also provide additional 
insight into the star-forming histories of the outer regions of these galaxies. 
Contamination from background galaxies and foreground 
stars, which our data demonstrates is already 
significant at the brightness of the RGB-tip when R$_{GC}^{LOS} 
= 9$ kpc in M33, will become an even greater issue at larger radii, and 
spectra will become increasingly important for identifying stars that belong to 
either galaxy. While spectroscopic observations of RGB-tip stars in Local Group 
galaxies are feasible with existing 8 -- 10 metre telescopes, studies of more 
distant systems, like NGC 2403, must await 20 -- 30 metre facilities. 
Nevertheless, infrared colors and narrow-band photometry, which 
can be obtained for these stars using $8 - 10$ metre telescopes, 
offer one means of distinguishing between halo and contaminating 
stars in an efficient manner; not only are stars near the AGB-tip 
relatively bright in $JHK$, but the near-infrared SEDs of 
dwarfs and giants differ, so that foreground stars can be identified.

	The late-type Sculptor group galaxies NGC 55 and NGC 247 are 
potentially interesting targets for surveys of bright 
AGB stars at large R$_{GC}$. Like M33 and NGC 2403, NGC 55 and NGC 247 are Sc 
galaxies. NGC 55 and NGC 247 have distances that are intermediate between 
M33 and NGC 2403, and are located near the South Galactic Pole, so that there 
will be lower contamination from foreground stars than in either NGC 2403 
or M33. NGC 55 is also viewed almost edge-on, thereby simplifying the task 
of studying the spatial distribution of stars away from the plane of the disk.

\acknowledgements{It is a pleasure to thank Sidney van den Bergh, Harvey 
Richer, and Jean-Rene Roy for commenting on an earlier version of the 
manuscript. Sidney van den Bergh also suggested 
that the globular cluster systems of M33 and NGC 2403 be considered when 
discussing the stellar contents of these galaxies. Thanks are also extended to 
Inger J{\o}rgensen, who co-ordinated the GMOS SV process, and to an anonymous 
referee, who made a number of comments that improved the paper.}

\parindent=0.0cm

\clearpage

\begin{table*}
\begin{center}
\begin{tabular}{lcc}
\tableline\tableline
Galaxy & $\mu_0$ & E(B--V)\tablenotemark{a} \\
\tableline
NGC 2403 & $27.51 \pm 0.24$ \tablenotemark{b} & 0.046 (0.052) \\
M33 & $24.81^{+0.15}_{-0.11}$ \tablenotemark{c} & 0.057 (0.079) \\
\tableline
\end{tabular}
\end{center}
\tablenotetext{a}{From Schlegel, Finkbeiner, \& Davis (1998). The entry in 
brackets is the reddening of the control field for that galaxy.}
\tablenotetext{b}{Freedman \& Madore (1988)}
\tablenotetext{c}{Kim et al. (2002)}
\caption{Adopted Distances and Reddenings}
\end{table*}

\clearpage

\begin{table*}
\begin{center}
\begin{tabular}{lcccccc}
\tableline\tableline
Field & RA & Dec & Filter & Exposure Time & FWHM \\
 & (J2000) & (J2000) & & (seconds) & (arcsec) \\
\tableline
M33 & 01:36:27.6 & $+$30:16:41 & $g'$ & $6 \times 60$ & 1.16 \\
 & & & $r'$ & $6 \times 60$ & 0.94 \\
 & & & $i'$ & $6 \times 60$ & 0.87 \\
 & & & $z'$ & $6 \times 60$ & 0.65 \\
 & & & & & \\
M33 control & 01:45:01.2 & $+$27:32:45 & $g'$ & $6 \times 60$ & 0.94 \\
 & & & $r'$ & $6 \times 60$ & 0.94 \\
 & & & $i'$ & $6 \times 60$ & 0.87 \\
 & & & $z'$ & $6 \times 60$ & 0.65 \\
 & & & & & \\
NGC 2403 & 07:37:51.1 & $+$65:43:06 & $g'$ & $6 \times 900$ & 1.23 \\
 & & & $r'$ & $6 \times 400$ & 0.87 \\
 & & & $i'$ & $6 \times 400$ & 0.80 \\
 & & & $z'$ & $12 \times 200$ & 0.72 \\
 & & & & & \\
NGC 2403 control & 07:52:32.6 & $+$61:23:32 & $g'$ & $2 \times 900$ & 1.02 \\
 & & & $r'$ & $6 \times 400$ & 0.65 \\
 & & & $i'$ & $6 \times 400$ & 0.65 \\
 & & & $z'$ & $12 \times 200$ & 0.87 \\
 & & & & & \\
\tableline
\end{tabular}
\end{center}
\caption{Details of the Observations}
\end{table*}

\clearpage

\begin{table*}
\begin{center}
\begin{tabular}{cc}
\tableline\tableline
Region & R$^{AGB}_{RGB}$ \\
\tableline
I$_g$ & $0.22 \pm 0.07$ \\
II$_g$ & $0.44 \pm 0.18$ \\
III$_g$ & $0.23 \pm 0.12$ \\
M33 & $0.9 \pm 0.5$ \\
\tableline
\end{tabular}
\end{center}
\caption{The ratio of bright AGB stars to upper RGB stars in each field}
\end{table*}

\clearpage

\begin{table*}
\begin{center}
\begin{tabular}{ccc}
\tableline\tableline
Field & Number of & Density \\
 & stars & (arcsec$^{-2}$) \\
\tableline
Region I$_g$ & $388.5 \pm 30.9$ & $(6.5 \pm 0.5) \times 10^{-6}$ \\
Region II$_g$ & $187.5 \pm 36.3$ & $(2.5 \pm 0.5) \times 10^{-6}$ \\
Region III$_g$ & $166.0 \pm 39.4$ & $(1.6 \pm 0.4) \times 10^{-6}$ \\
M33 & $31.0 \pm 12.7$ & $(1.3 \pm 0.5) \times 10^{-7}$ \\
M33$_{NGC2403}$ & -- & $(2.1 \pm 0.8) \times 10^{-6}$ \\
\tableline
\end{tabular}
\end{center}
\caption{The density of stars within 1.5 mag of the RGB-tip}
\end{table*}

\clearpage

\begin{table*}
\begin{center}
\begin{tabular}{lc}
\hline\hline
Field & Surface Brightness ($r-$band)\tablenotemark{a} \\
\hline
Region I$_g$ & 26.6 \\
Field 1 & 21.8 \\
Fields 2 $+$ 3 & 24.3 \\
\hline
\end{tabular}
\end{center}
\caption{$r-$band surface brightness measurements for Region I$_g$ and the Hudon et al. fields}
\tablenotetext{a}{From Kent (1987)}
\end{table*}

\clearpage

\begin{table*}
\begin{center}
\begin{tabular}{lcccc}
\tableline\tableline
Object & [Fe/H] & $(V-I)_{RGB}$ & $(R-I)_{RGB}$ & $(r'-i')_{RGB}$ \\
\tableline
N104 & --0.7 & 1.54 & 0.74 & 0.50 \\
N1851 & --1.3 & 1.38 & 0.65 & 0.41 \\
N6752 & --1.5 & 1.28 & 0.60 & 0.36 \\
N7089 & --1.6 & 1.26 & 0.59 & 0.35 \\
N7078 & --2.2 & 1.17 & 0.54 & 0.30 \\
 & & & & \\
And V & -2.2 & -- & -- & 0.36 \\
N2419 & -2.2 & -- & -- & 0.34 \\
\tableline
\end{tabular}
\end{center}
\caption{$r'-i'$ colors one mag below the RGB-tip for selected 
globular clusters and the dwarf spheroidal galaxy And V}
\end{table*}

\clearpage

\begin{center}
FIGURE CAPTIONS
\end{center}

\figcaption
[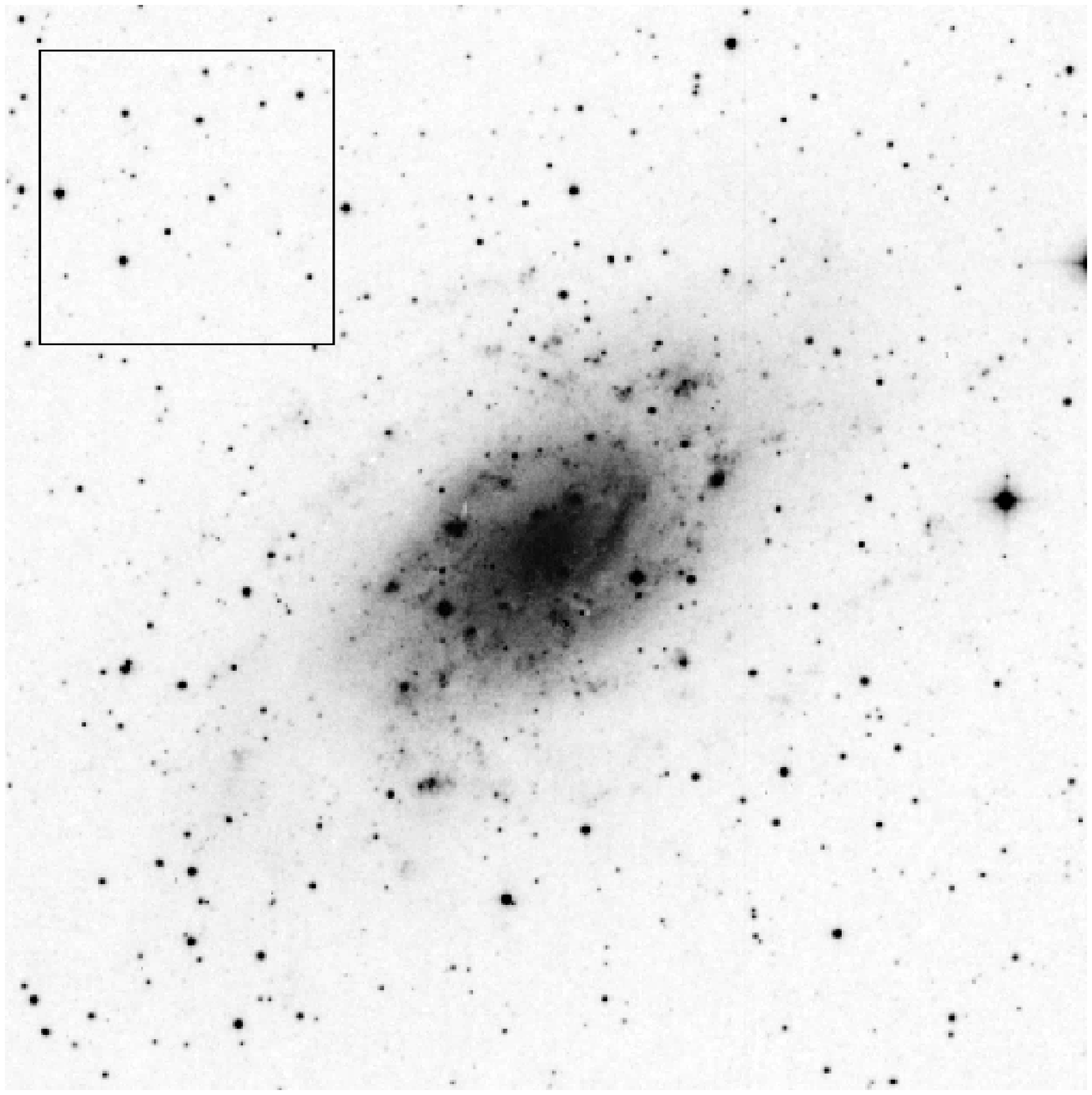]
{A $20 \times 20$ arcmin section of the Digital Sky Survey (DSS) centered on 
NGC 2403. The box shows the location of the $5.5 \times 5.5$ arcmin 
GMOS field. Although not obvious from this figure, the lower right hand 
corner of this field contains bright young main sequence 
stars belonging to the disk of NGC 2403. North is at the top, and East 
is to the left.}

\figcaption
[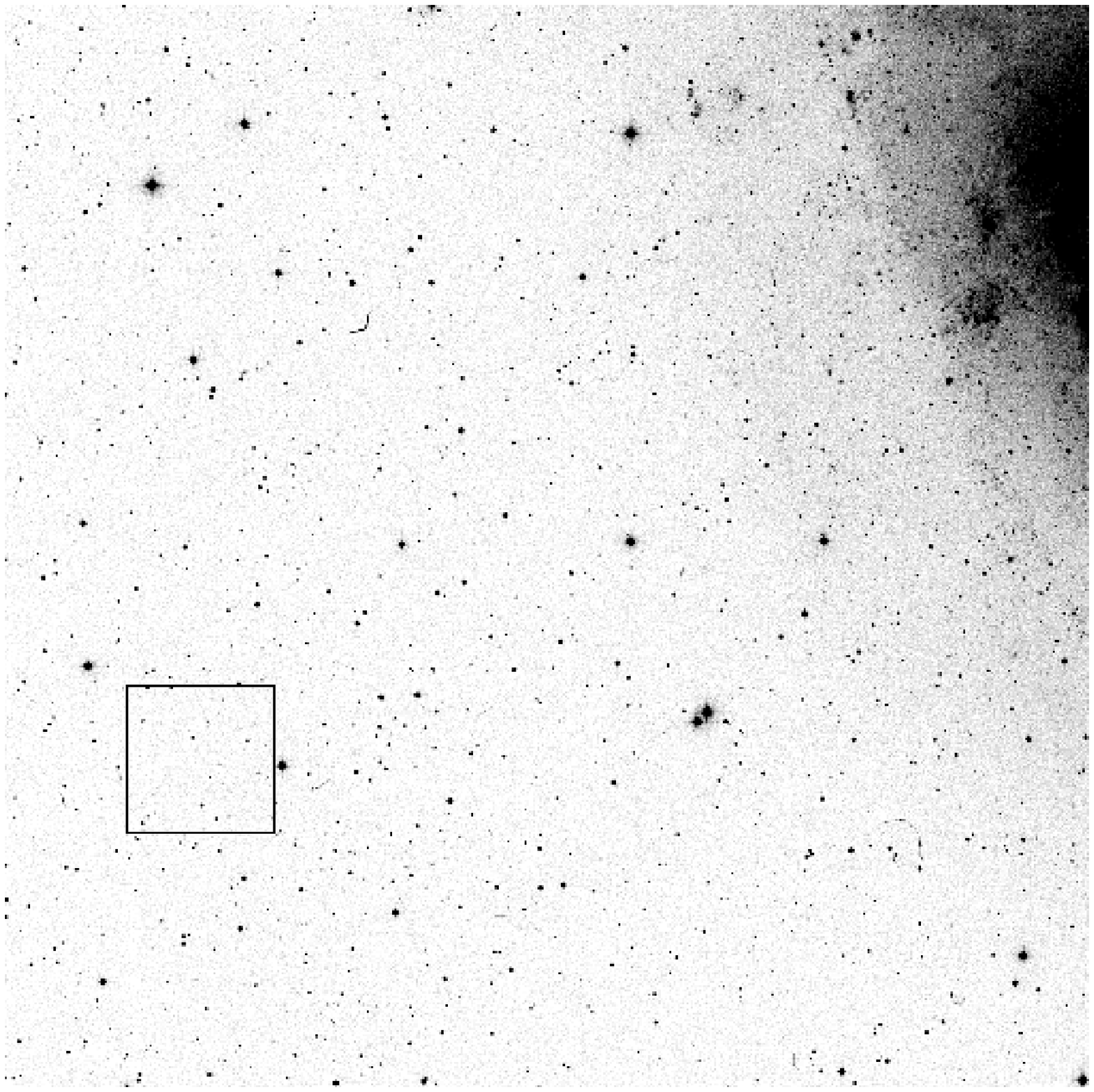]
{A $40 \times 40$ arcmin section of the DSS near M33. The box shows the 
location of the $5.5 \times 5.5$ arcmin GMOS field. North is at the top, and 
East is to the left. The area observed by MK86 is near the center of the 
figure.} 

\figcaption
[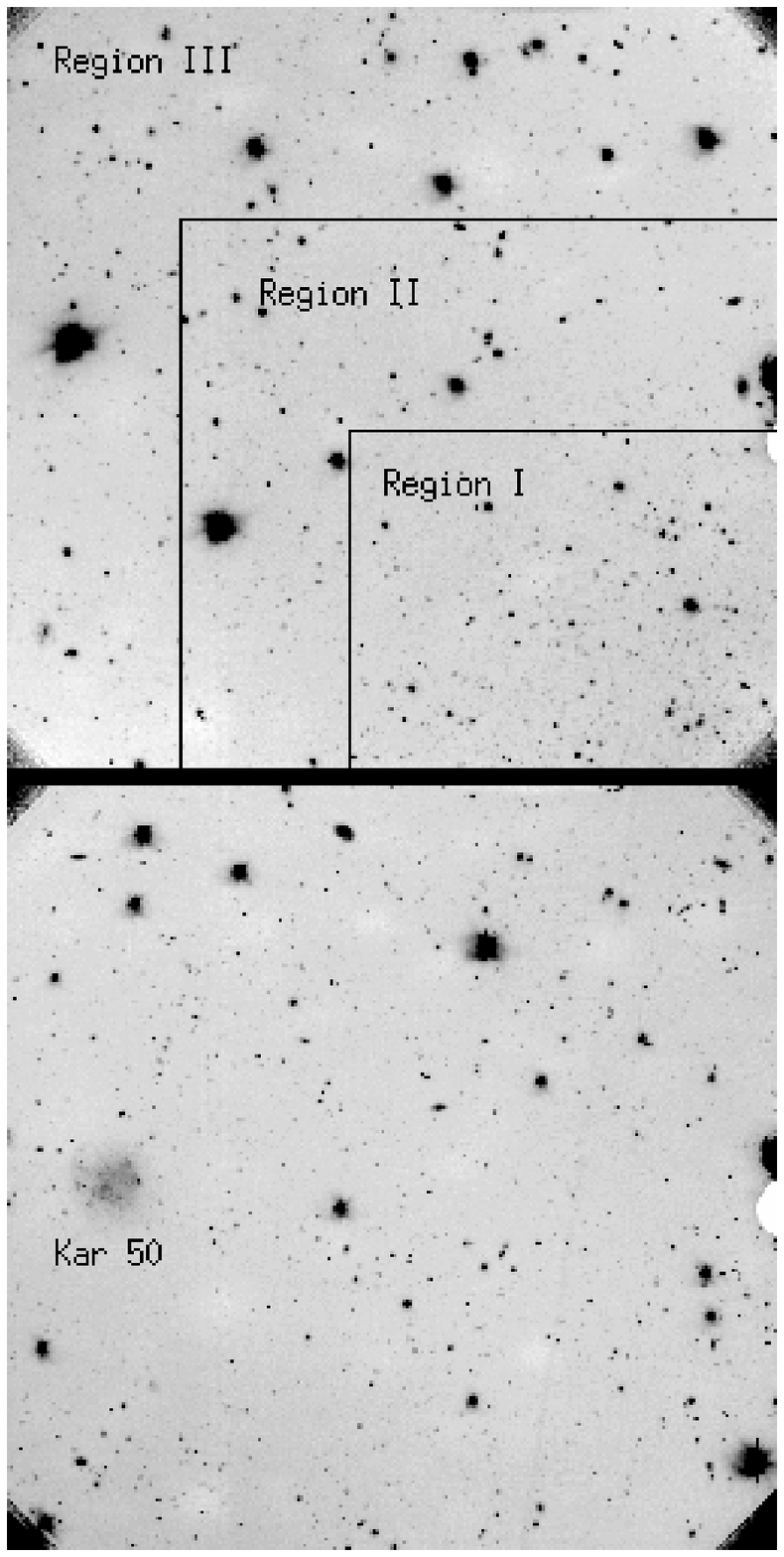]
{The $i'$ images of the NGC 2403 galaxy (top panel) and control (bottom panel) 
fields. Each image covers approximately $5.5 \times 5.5$ arcmin, with North at 
the top and East to the left. The three regions used to analyze 
the stellar content of this field are indicated. The dwarf 
irregular galaxy Kar 50, which was discussed by Davidge (2002), is labelled in 
the lower panel.}

\figcaption
[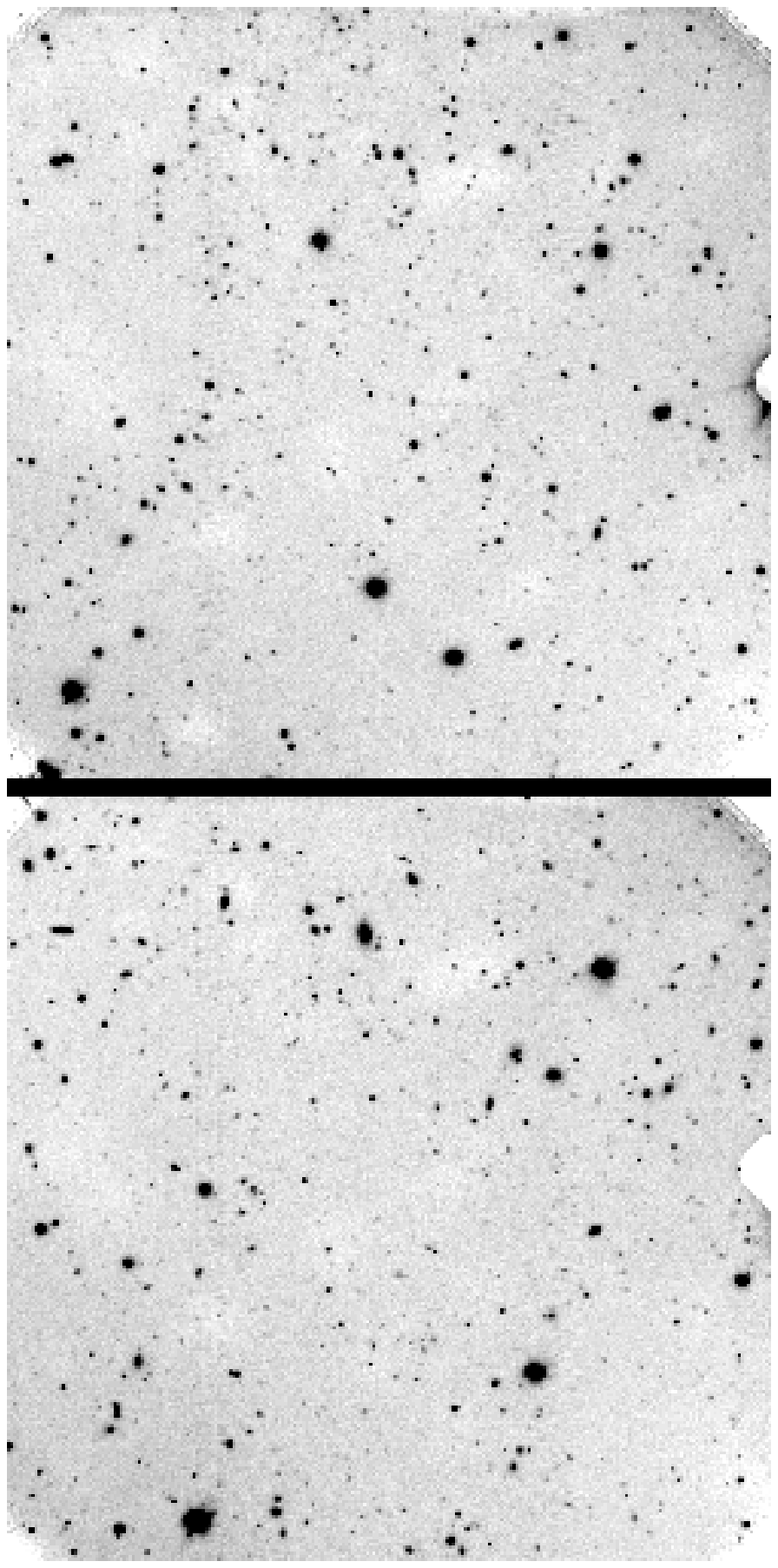]
{The $i'$ images of the M33 galaxy (top panel) and control (bottom panel) 
fields. Each image covers approximately $5.5 \times 5.5$ arcmin, with 
North at the top and East to the left.}

\figcaption
[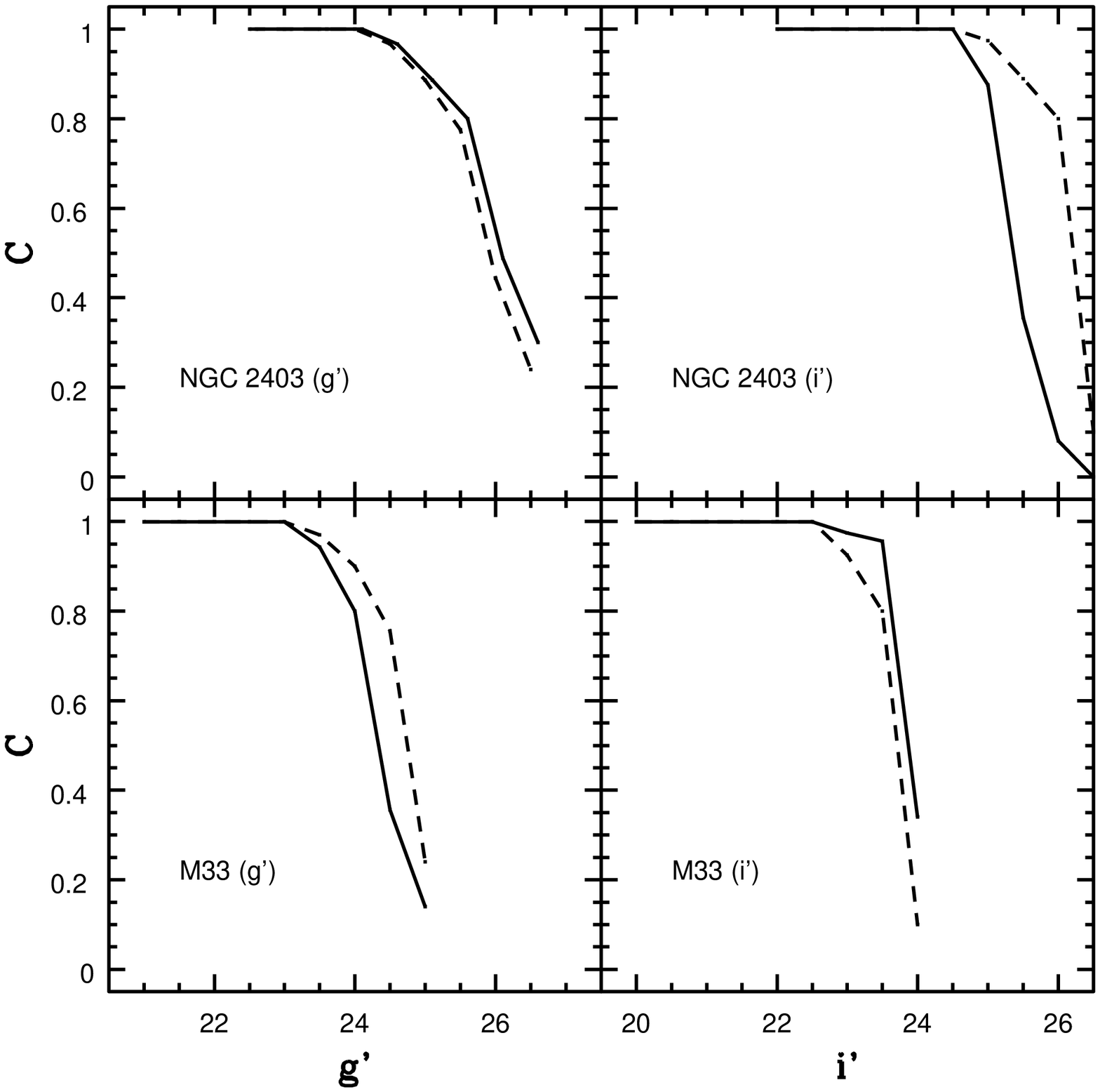]
{The completeness fractions $C$ in $g'$ and $i'$, 
as computed from artificial star experiments. $C$ is the ratio of 
detected artificial stars to the total number inserted into each data frame 
per 0.5 mag interval. The completeness curves for the outer 
regions of NGC 2403 and M33 are shown as solid lines, while the corresponding 
curves for the control fields are shown as dashed lines. The NGC 2403 data 
go deeper than the M33 data because of the difference in total exposure time 
(Table 1).} 

\figcaption
[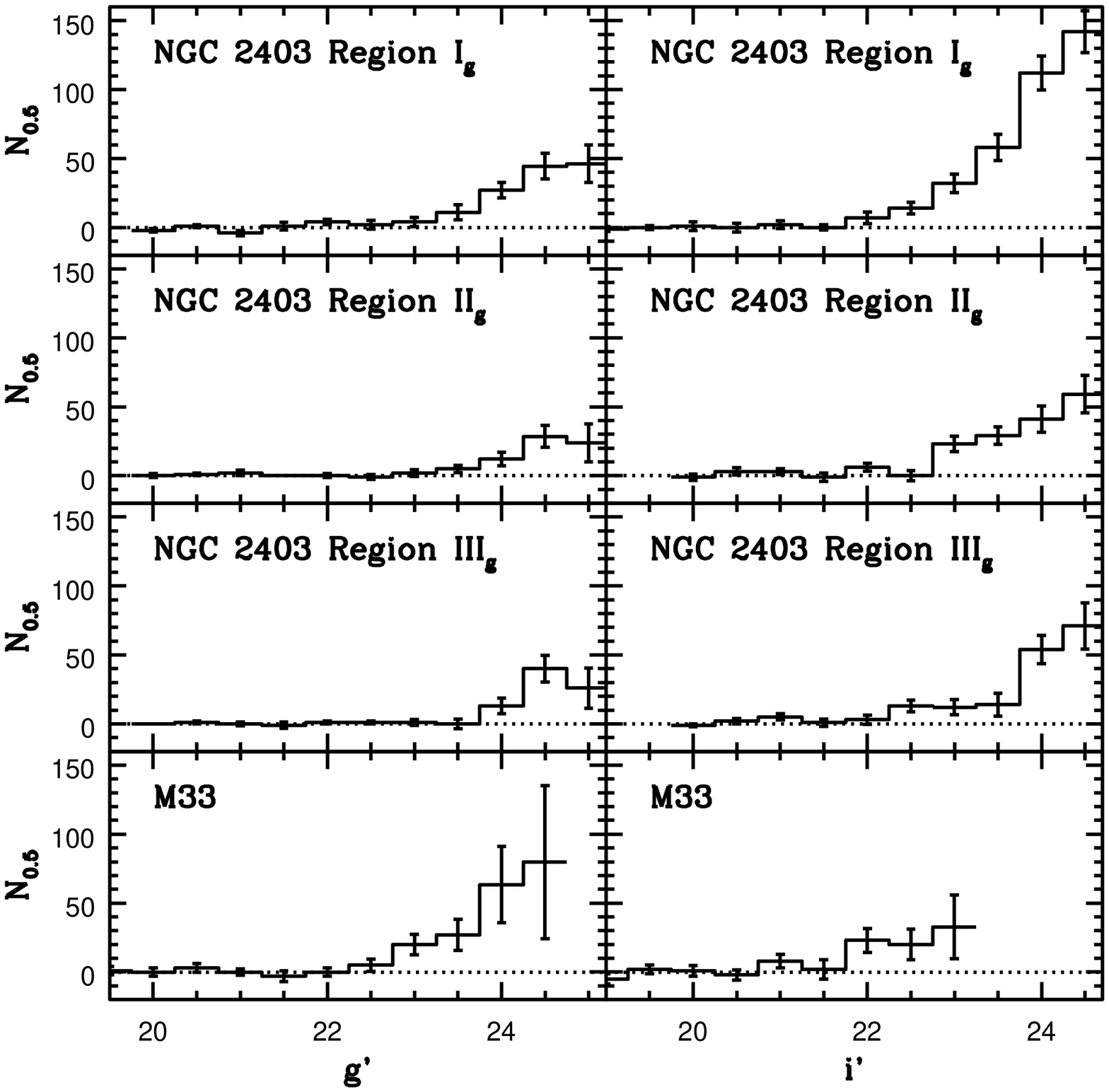]
{The $g'$ and $i'$ LFs of sources in the outer regions of NGC 2403 and M33. 
N$_{0.5}$ is the difference between the number of sources per 0.5 mag interval 
in the galaxy and control fields, corrected for incompleteness; 
the dotted line indicates N$_{0.5} = 0$. 
The $g'$ LFs show sources detected in both $g'$ and $r'$, while the 
$i'$ LFs show sources detected in both $r'$ and $i'$. The error bars reflect 
uncertainties due to Poisson statistics and the completeness corrections.} 

\figcaption
[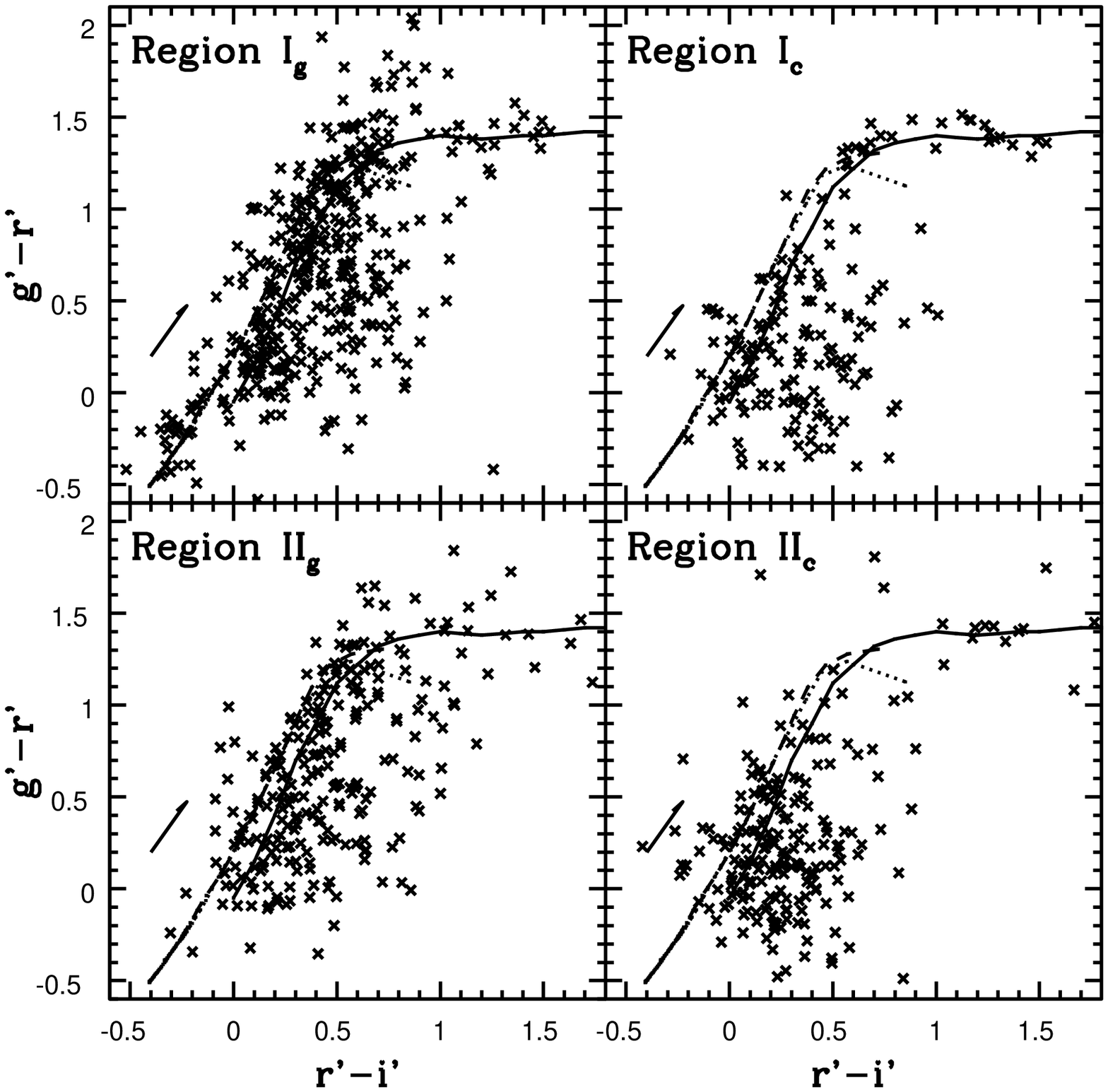]
{The ($g'-r', r'-i'$) TCDs of NGC 2403 Regions I and II in the galaxy and 
control fields. The dotted and dashed lines show the loci of the log(g)=2.5 
and 4.5 solar metallicity models of Lenz et al. (1998) while the solid line is 
the stellar sequence from Figure 3 of Schneider et al. (2001). The reddening 
vector, which has a length corresponding to A$_{g'} = 1$ mag, follows the 
relations listed in Table 6 of Schlegel et al. (1998), which in turn are taken 
from the Cardelli, Clayton, \& Mathis (1989) and O'Donnell (1994) reddening 
curves. Note that there are a number of sources in Regions I$_g$ and II$_g$ 
with $r'-i' > 0.4$ and $g'-r' > 1.5$, and objects with corresponding colors are 
largely abscent in Regions I$_c$ and II$_c$. These red objects are interpreted 
as giants that are subject to reddening internal to NGC 2403.}

\figcaption
[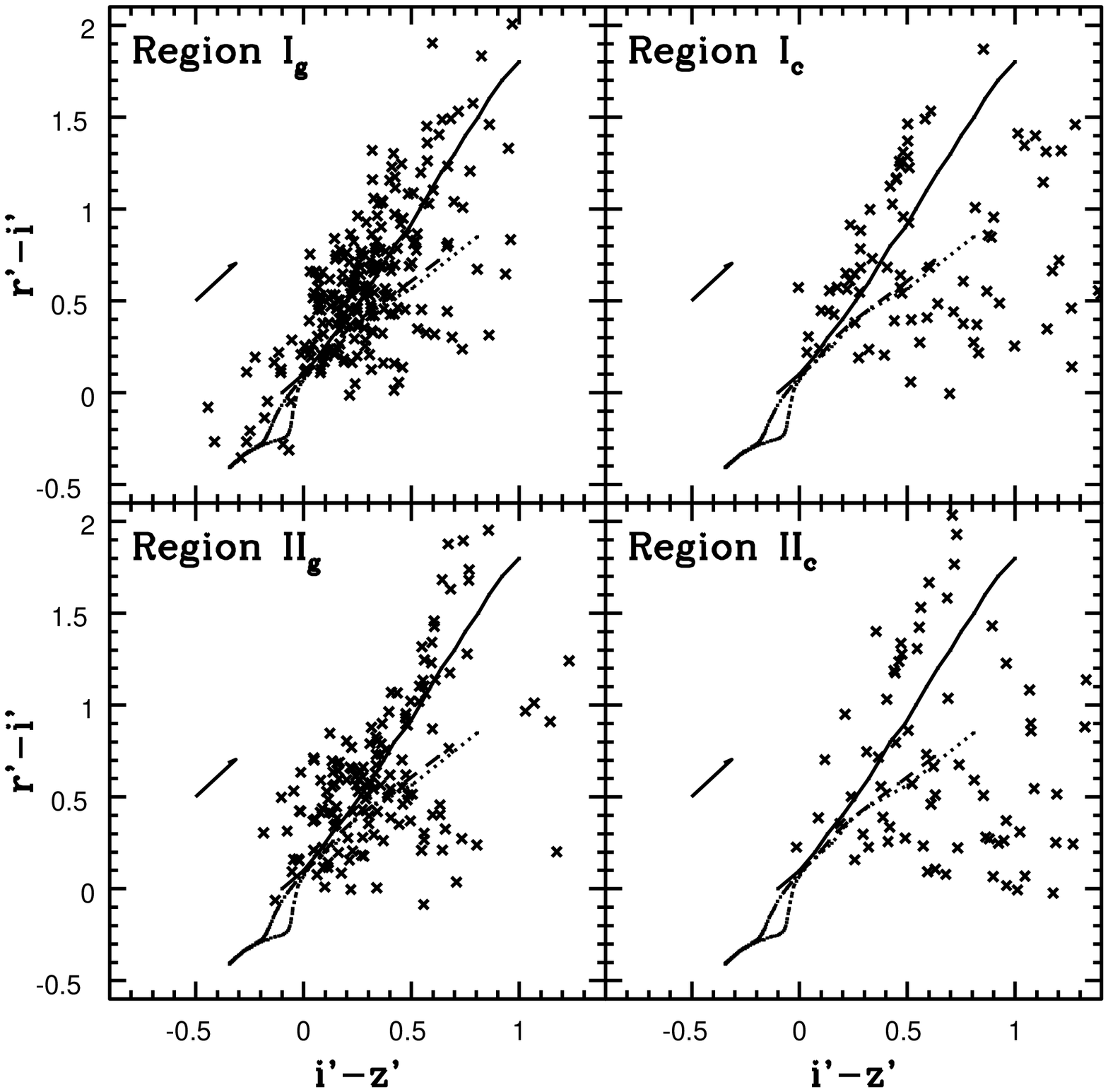]
{The ($r'-i', i'-z'$) TCDs of NGC 2403 Regions I and II in the galaxy and 
control fields. The dotted and dashed lines show the loci of the log(g)=2.5 and 
4.5 solar metallicity models of Lenz et al. (1998) while the solid line is the 
stellar sequence from Figure 3 of Schneider et al. (2001). 
The reddening vector, which has a length 
corresponding to A$_{g'} = 1$ mag, follows the relations 
listed in Table 6 of Schlegel et al. (1998), which in turn are taken from the 
Cardelli et al. (1989) and O'Donnell (1994) reddening curves.}

\figcaption
[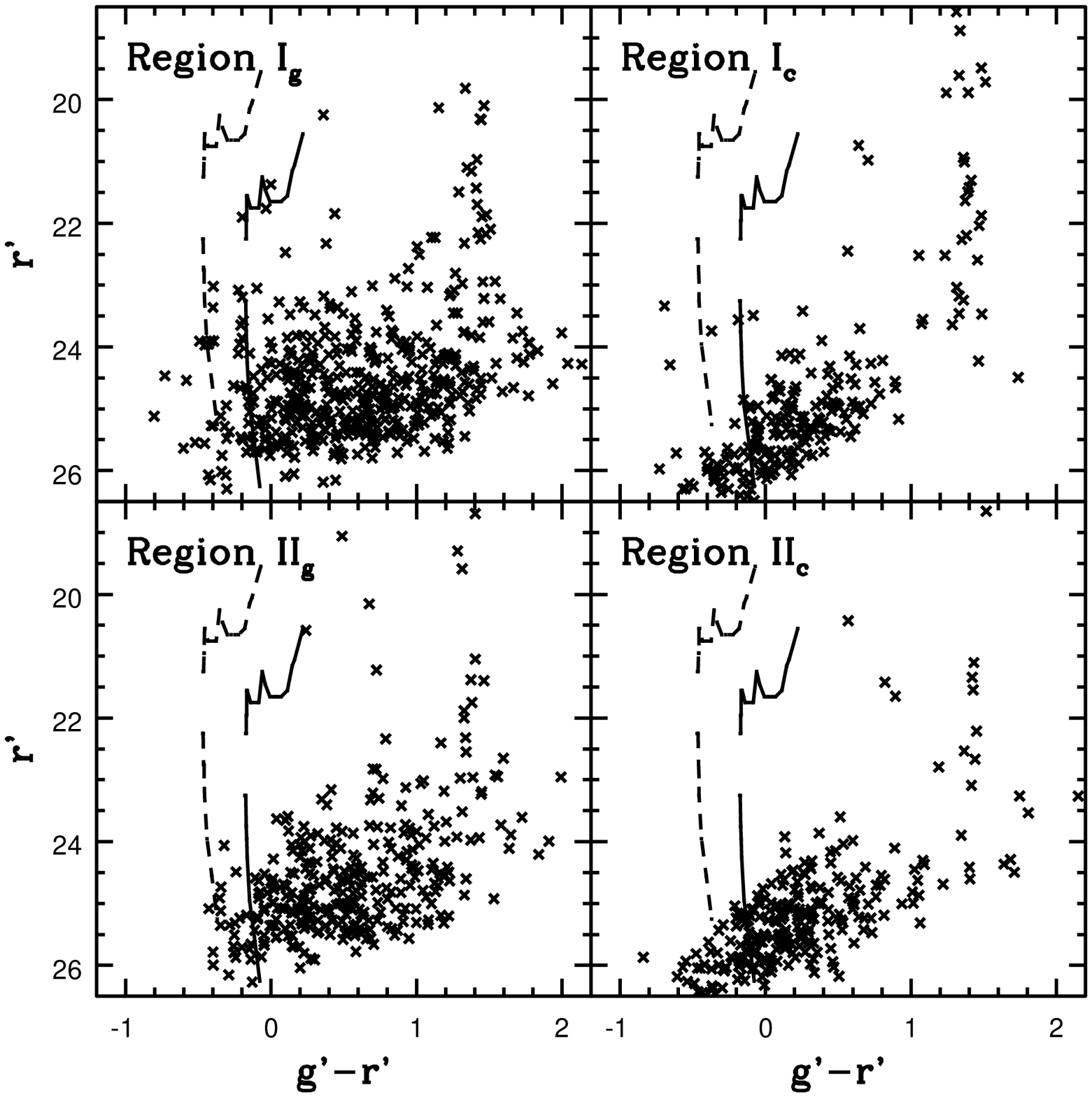]
{The $(r', g'-r')$ CMDs of NGC 2403 Regions I and II in the galaxy and control 
fields. The dashed lines in the left hand panel are the main sequence and 
supergiant Ia sequence defined by stars in the Galaxy and the LMC, shifted to 
match the distance of NGC 2403 with a correction for foreground reddening only.
The solid lines are the same sequences, but with an assumed internal extinction 
of A$_{g'} = 1$ mag for NGC 2403.}

\figcaption
[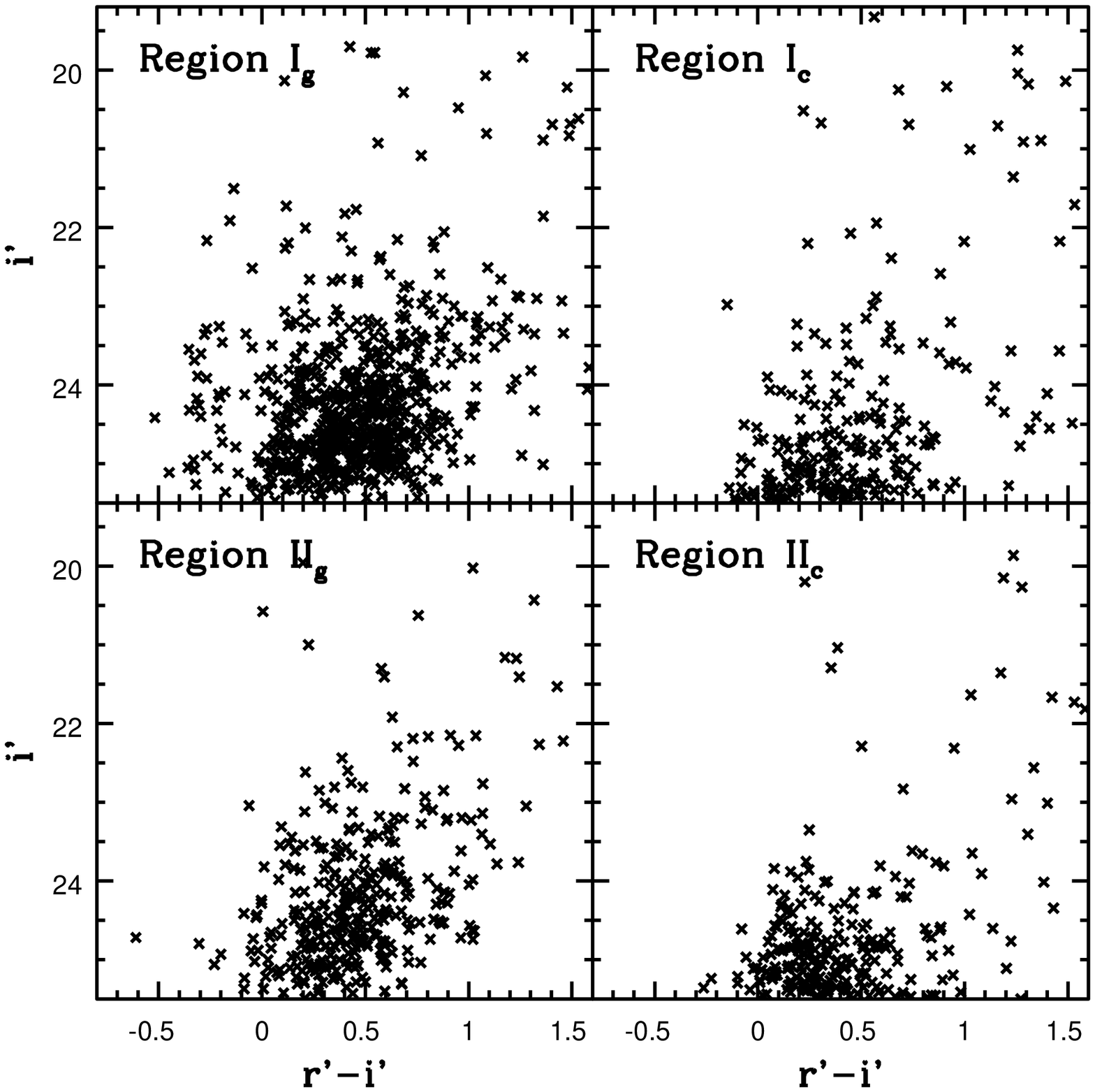]
{The $(i', r'-i')$ CMDs of NGC 2403 Regions I and II in the galaxy and control 
fields. Both Regions I$_g$ and II$_g$ contain an excess population of sources 
with respect to the corresponding control fields when $i' > 23.5$, due to stars 
evolving on the RGB in NGC 2403.}

\figcaption
[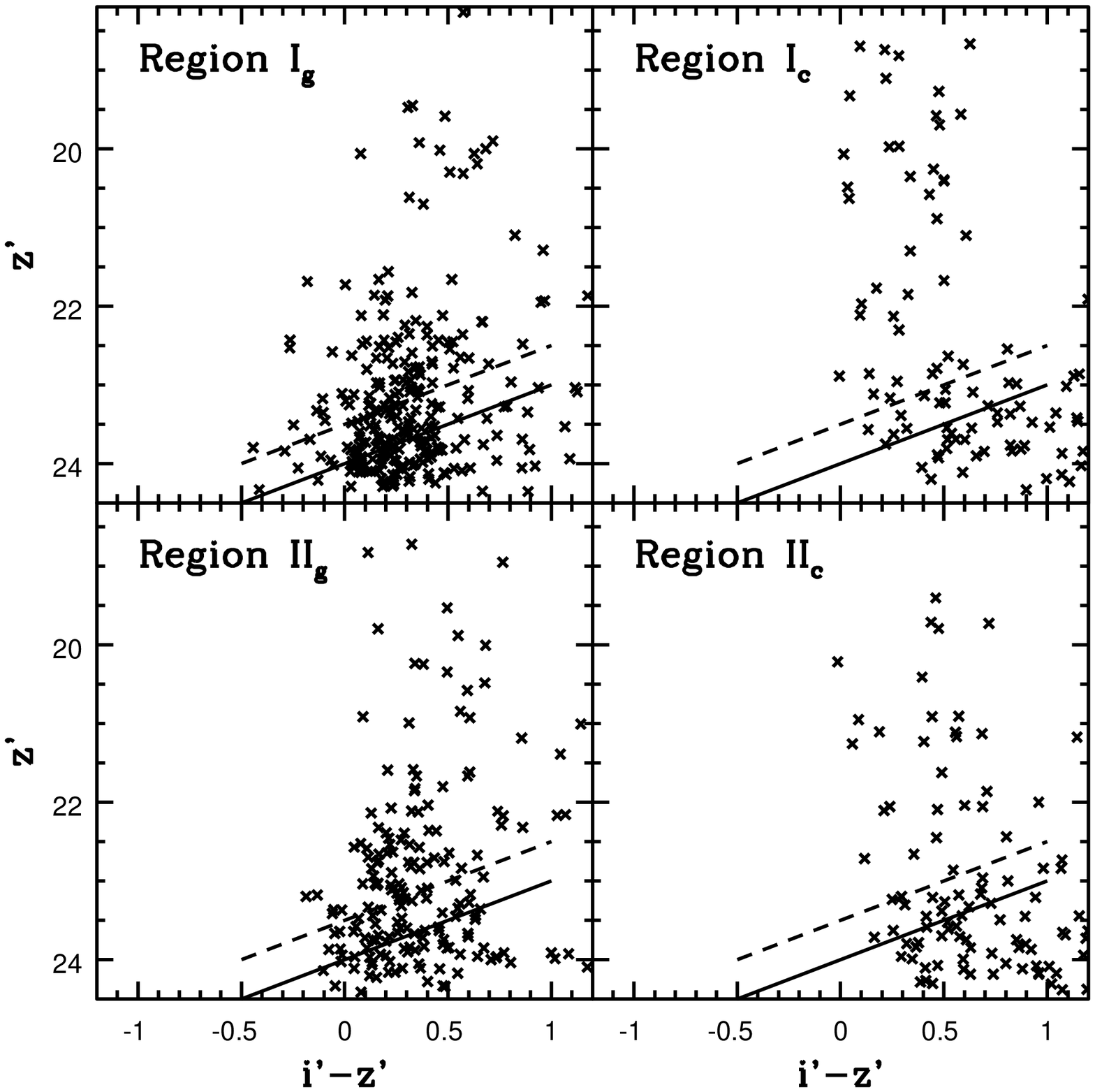]
{The $(z', i'-z')$ CMDs of NGC 2403 Regions I and II 
in the galaxy and control fields. The dashed line shows $i' = 23.5$, which 
corresponds to the approximate brightness of the RGB-tip with no internal 
extinction assumed for NGC 2403, while the solid line shows $i' = 24$, 
which corresponds to the RGB-tip with an assumed internal extinction A$_{g'} = 
1$ mag. Regions I$_g$ and II$_g$ contain an excess population of sources with 
respect to the corresponding control fields when $z' > 
21.5$, and the majority of these are AGB stars in NGC 2403. 
Aside from a difference in the number of stars, the overall 
morphologies of the $(z', i'-z')$ CMDs of Region I$_g$ and II$_g$ are not 
greatly different, suggesting similar AGB contents.}

\figcaption
[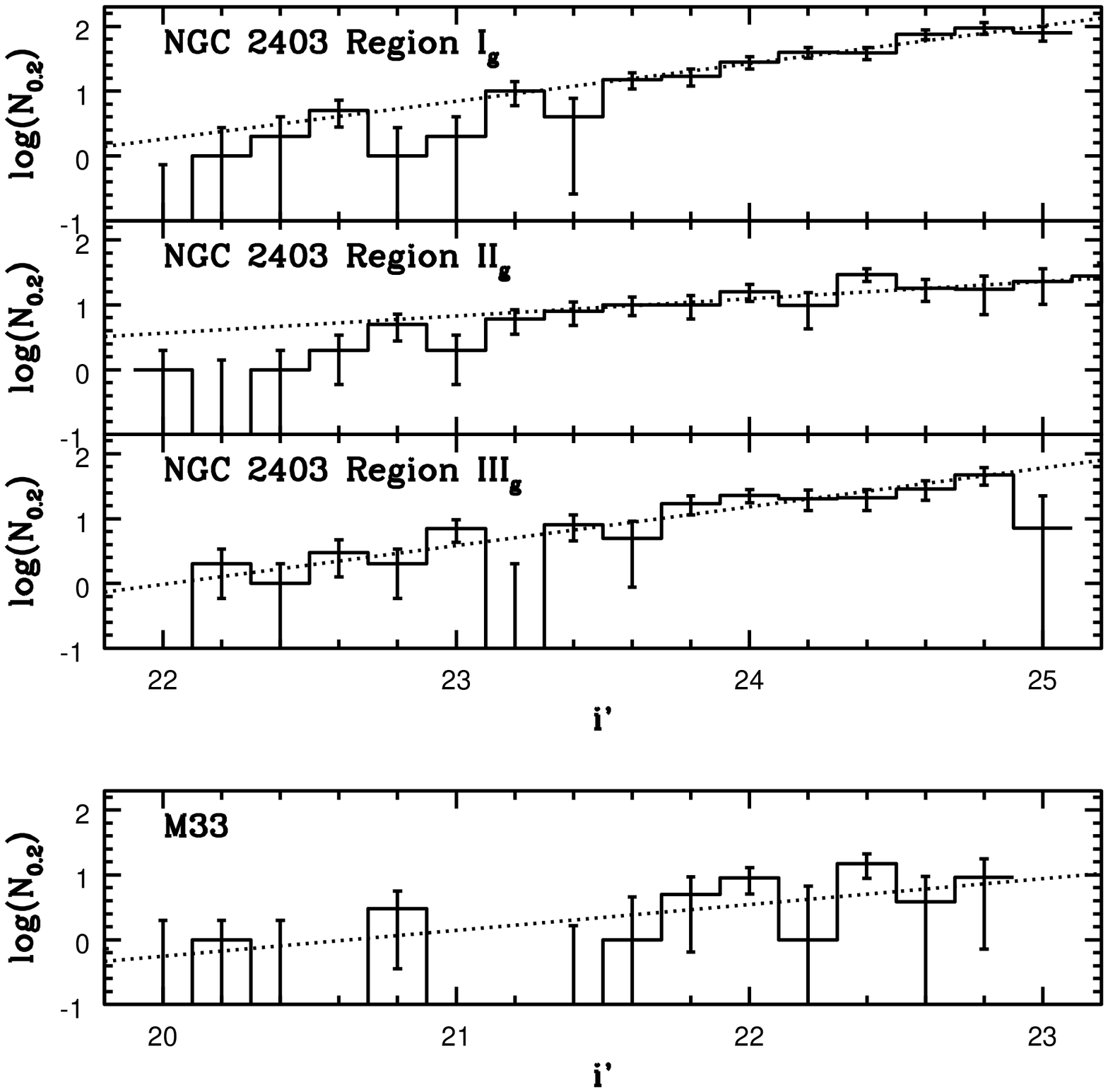]
{The $i'$ LFs of RGB stars in the outer regions of NGC 2403 and M33. N$_{0.2}$ 
is the difference between the number of stars per 0.2 mag $i'$ interval 
in the galaxy and control fields in these color intervals, corrected 
for incompleteness. The errorbars reflect the uncertainties due to 
Poisson statistics and the completeness corrections. The dotted lines 
in the top three panels show power laws that have been fit to the 
data in each NGC 2403 Region when $i' > 23.6$, while the dotted line in the 
bottom panel shows a power-law with an exponent of 0.4 that 
has been fit to the M33 LF when $i'$ is between 21.6 and 22.8.}

\figcaption
[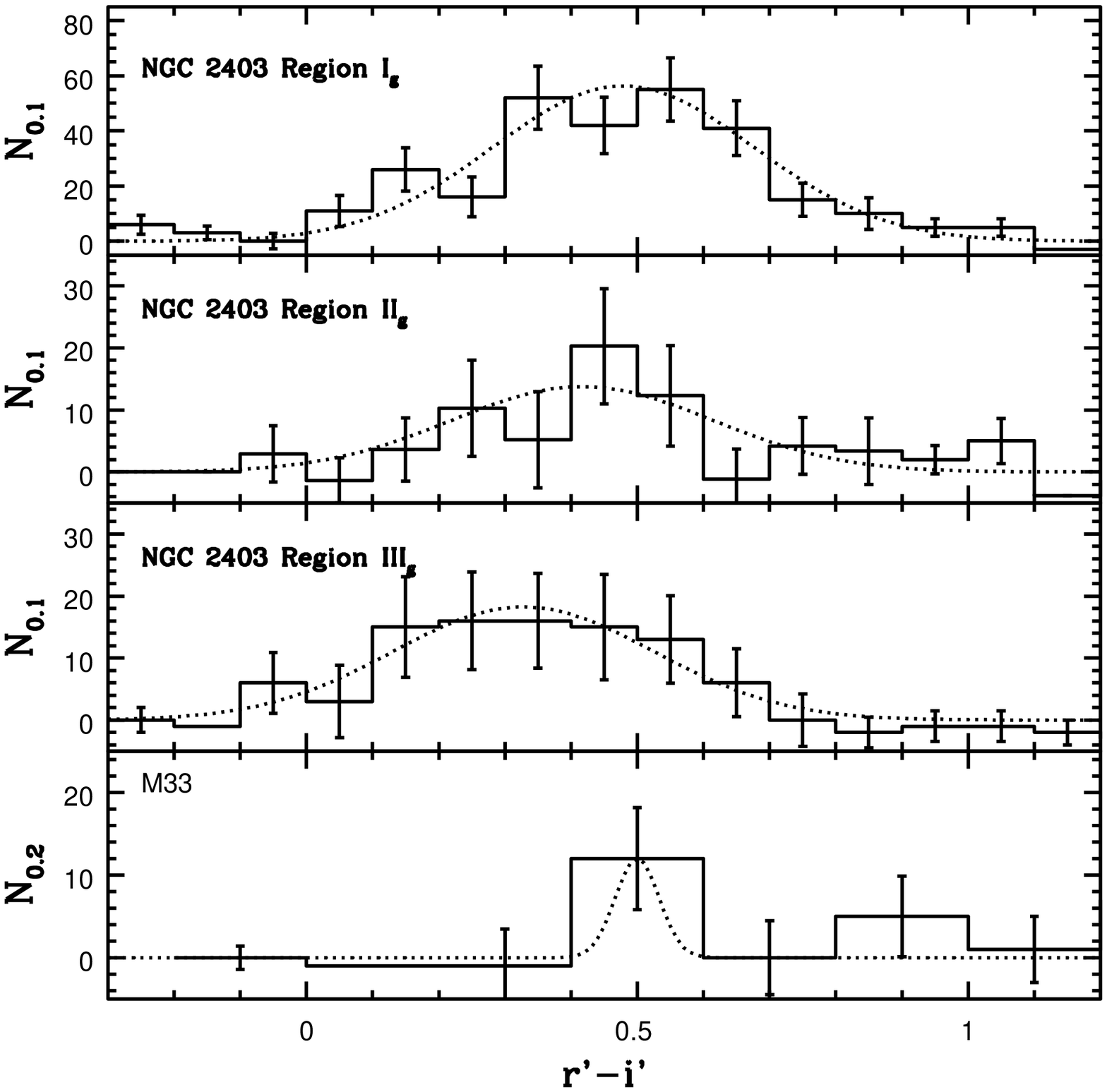]
{The $r'-i'$ distributions of sources on the upper RGB in NGC 2403 and M33. 
The $r'-i'$ distributions for NGC 2403 were computed using stars with 
$i'$ between 24 and 25, while the distribution for M33 uses stars with $i'$ 
between 21.3 and 22.3. The solid lines show the difference between the color 
distributions in the galaxy and control fields in these brightness intervals, 
while the error bars show the uncertainties due to Poisson statistics. 
N$_{0.1}$ and N$_{0.2}$ are the numbers of stars per 0.1 mag and 0.2 
mag intervals in $r'-i'$. The dotted lines in the top three panels are 
gaussians with $\sigma = \pm 0.197$ mag, which is the photometric error 
predicted by the artificial star experiments, centered on $\overline{r'-i'}$ 
and scaled to match the number of sources with $r'-i'$ between 0.3 and 0.7 in 
each region in NGC 2403. Note the excellent 
agreement with the observed distributions. The dotted line in the bottom panel 
is a gaussian with $\sigma = \pm 0.03$ mag, which is the photometric 
error predicted by the artificial star experiments for the M33 data, 
and $\overline{r'-i'} = 0.5$. The gaussian in the bottom panel has been scaled 
to match the number of sources in the $r'-i' = 0.5$ bin.}

\figcaption
[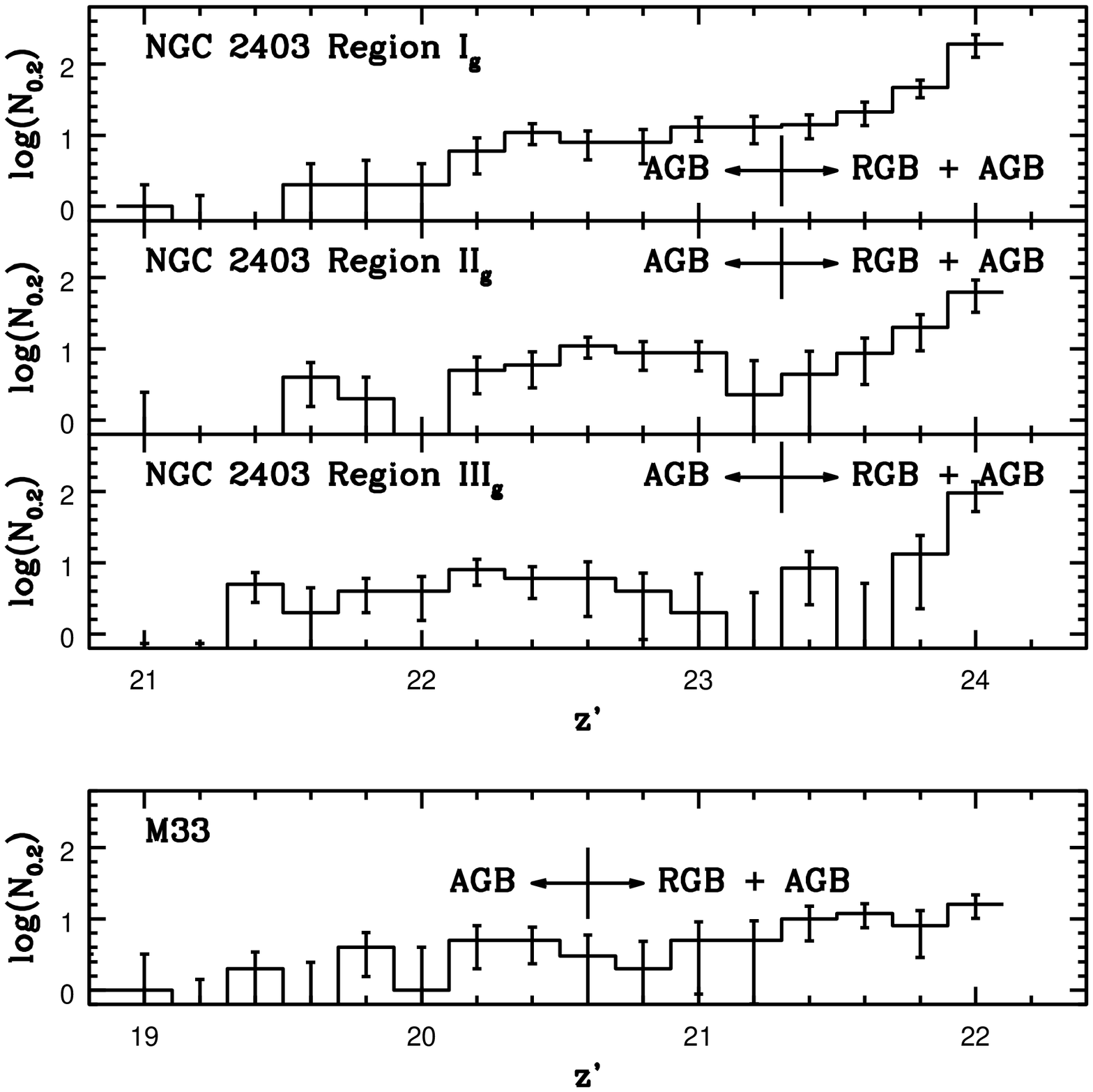]
{The $z'$ LFs of stars with $i'-z'$ between 0 and 1 in the outer regions of 
NGC 2403 and M33. N$_{0.2}$ is the difference between the number of stars per 
0.2 mag $z'$ interval in the galaxy and control fields, corrected 
for incompleteness. The errorbars reflect the uncertainties due to 
Poisson statistics and the completeness corrections. The approximate 
brightness of the RGB-tip is indicated with a vertical line in each panel; 
sources fainter than the RGB-tip are a combination of RGB and AGB stars, while 
sources brighter than the RGB-tip are on the AGB. Note that there is a 
significant number of bright AGB stars in each field.}

\figcaption
[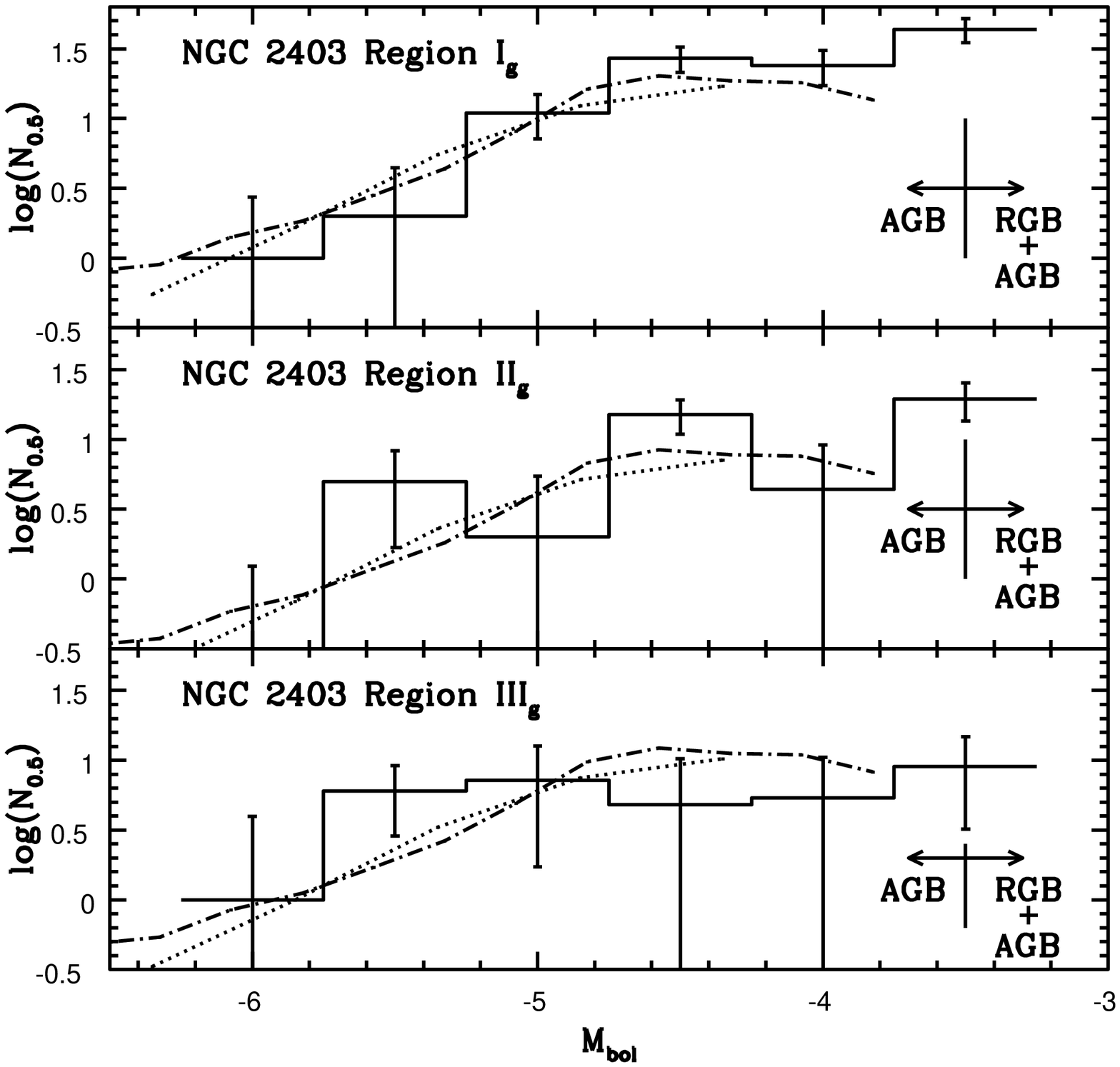]
{The bolometric LFs of AGB stars in the outer regions of NGC 2403. N$_{0.5}$ 
is the difference between the number of stars per 0.5 mag M$_{bol}$ interval
in the galaxy and control fields, corrected 
for incompleteness. The errorbars include the uncertainties due to 
Poisson statistics and the completeness corrections. The LFs were 
constructed by first transforming $i'$ and $r'-i'$ into I$_{KC}$ 
and $(R-I)_{KC}$ using the procedure described in the text, and then applying 
the criteria for identifying AGB stars discussed 
by Hudon et al. (1989). Bolometric corrections were computed from 
Equation 2 of Bessell \& Wood (1984). The dotted line in each 
panel is the AGB LF of Hudon et al. (1989) NGC 2403 Fields 2 and 3, while 
the dashed-dotted line is the LF of AGB stars in the LMC from Reid \& Mould 
(1984). The Hudon et al. and Reid \& Mould LFs have been scaled to match the 
number of objects in each region of NGC 2403 with M$_{bol}$ between --6.2 and 
--4.5, while the Hudon et al. (1989) LF has also been shifted by 0.1 mag to 
match the distance modulus adopted in the current study. Note that both the 
Hudon et al. and Reid \& Mould LFs match the current observations 
within the estimated uncertainties. The approximate magnitude of the 
RGB-tip is indicated with the vertical line in each panel; 
sources fainter than the RGB-tip are a combination of RGB and AGB stars, while 
sources brighter than the RGB-tip are on the AGB.} 

\figcaption
[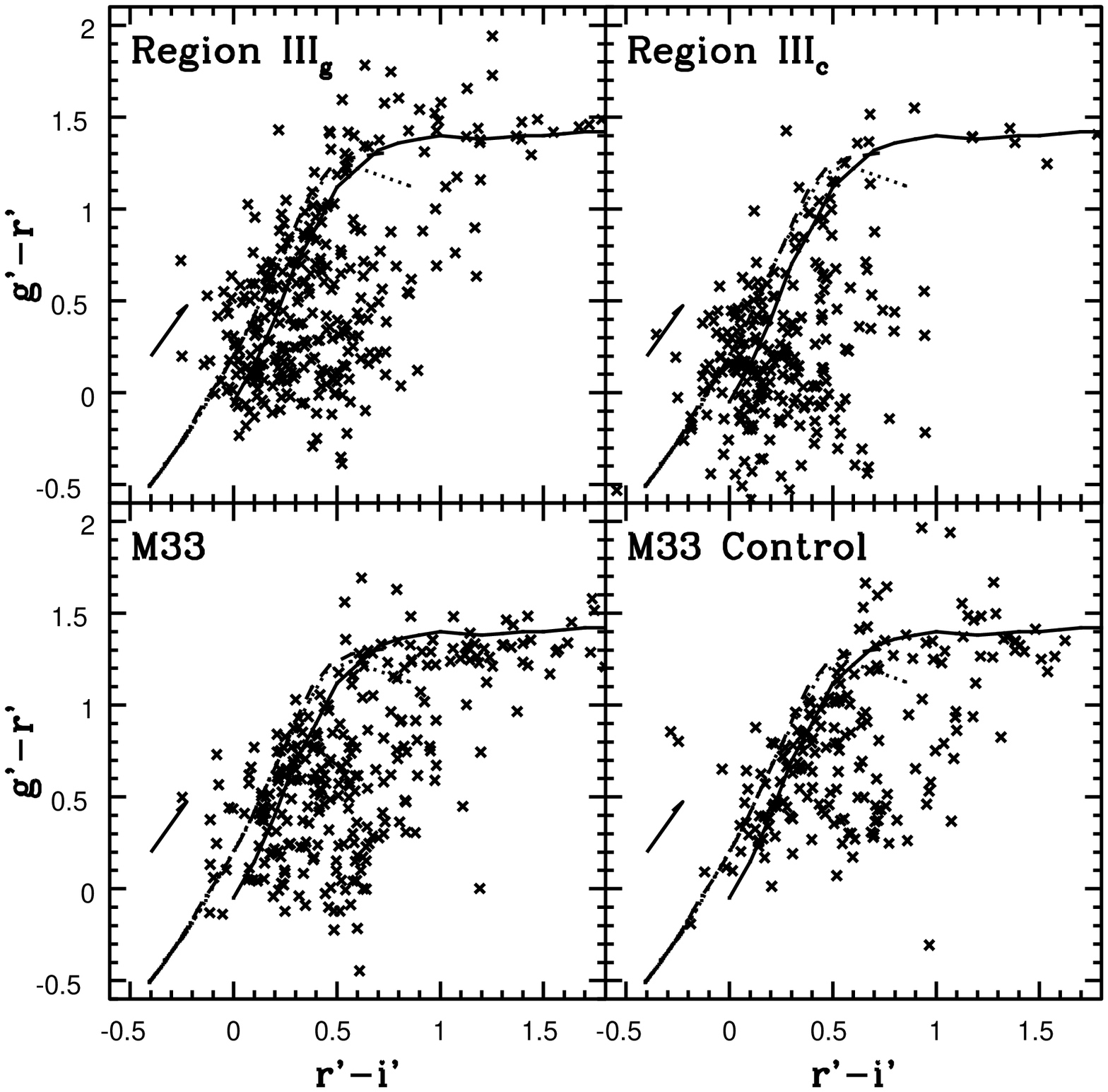]
{The ($g'-r', r'-i'$) TCDs of the galaxy and control fields for NGC 2403 
Region III and M33. See Figure 7 for additional details.}

\figcaption
[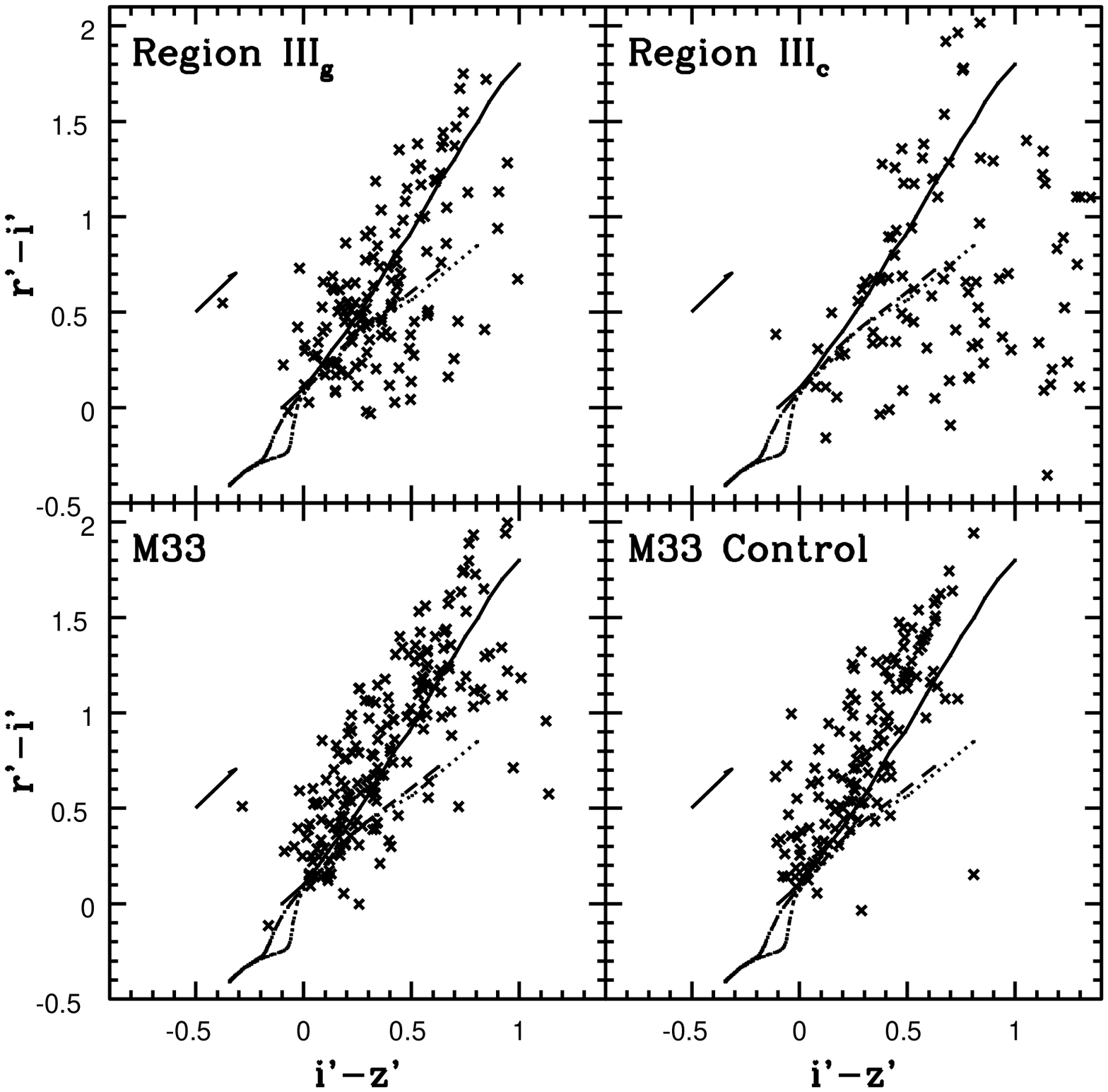]
{The ($r'-i', i'-z'$) TCDs of the galaxy and control fields for NGC 2403 
Region III and M33. See Figure 8 for additional details.}

\figcaption
[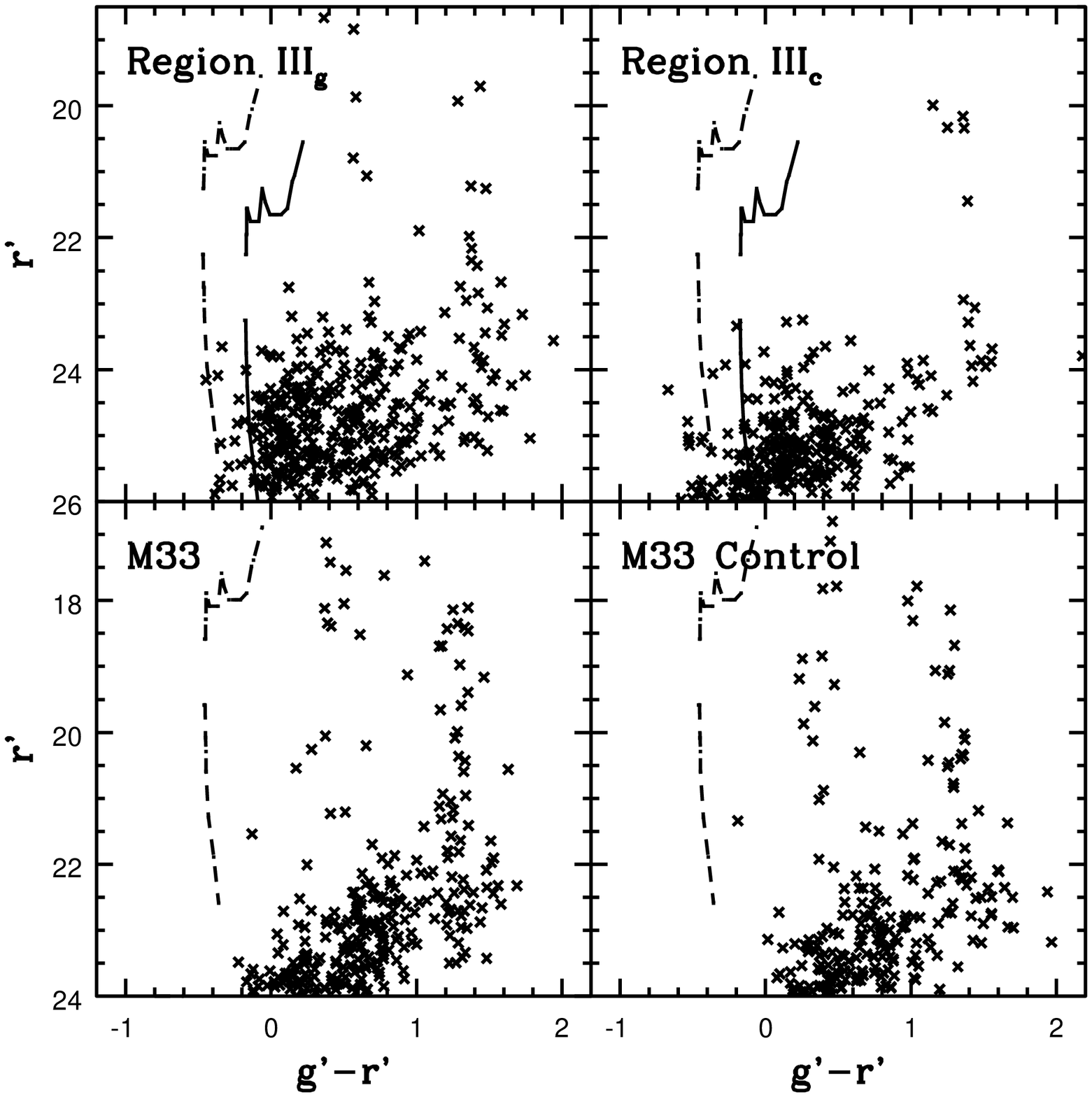]
{The $(r', g'-r')$ CMDs of the galaxy and control fields for NGC 2403 
Region III and M33. The dashed lines in the upper panels are the main 
sequence and supergiant Ia sequence defined by stars in the Galaxy and the LMC, 
shifted to match the distance of NGC 2403 for two reddenings (foreground only, 
and foreground with an internal extinction A$_{g'} = 1$ mag). The dashed lines 
in the lower panel are the main sequence and supergiant Ia sequences shifted to 
the distance of M33, with only foreground reddening. See Figure 9 for 
additional details.} 

\figcaption
[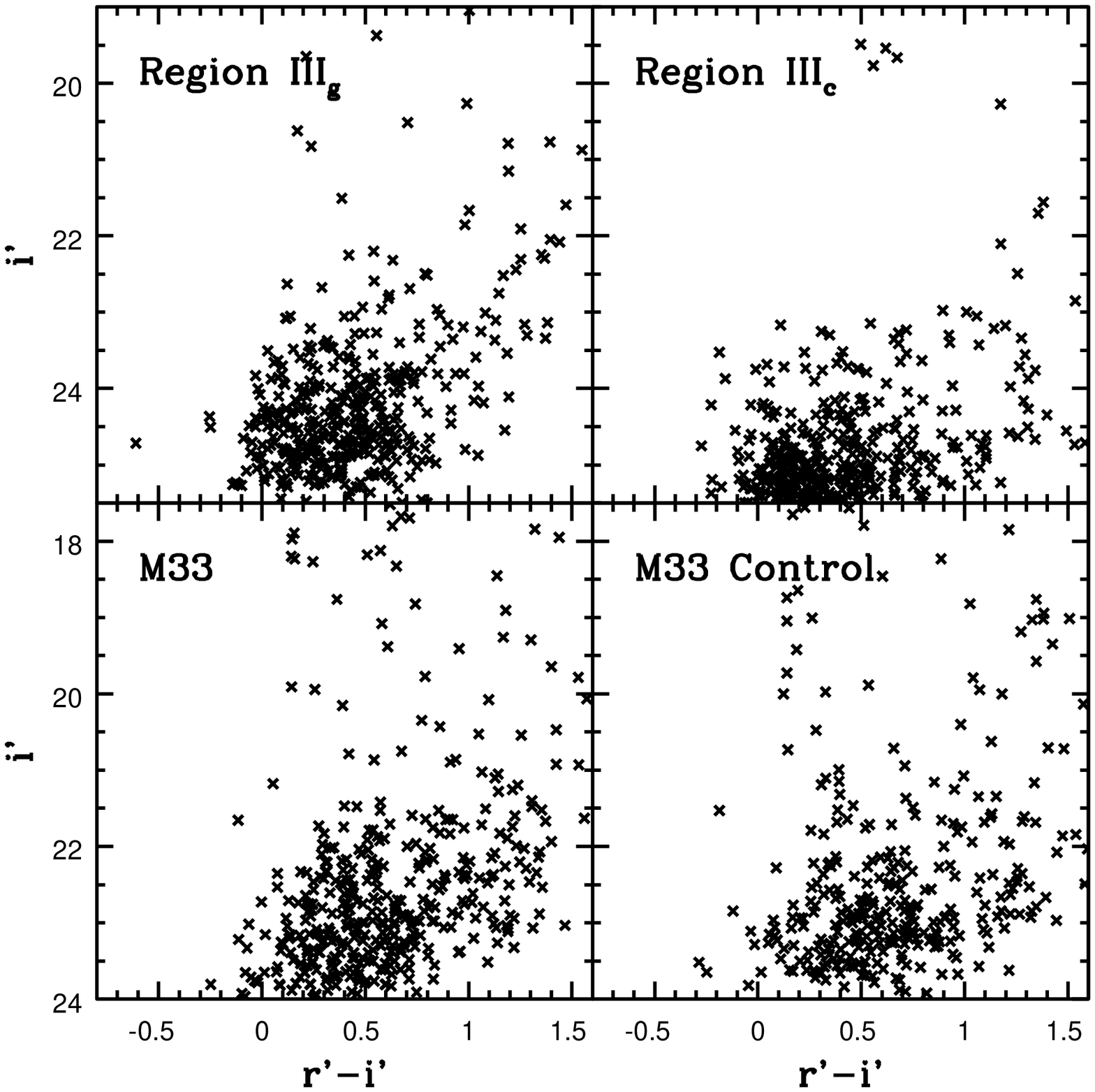]
{The $(i', r'-i')$ CMDs of the galaxy and control fields for NGC 2403 Region 
III and M33. The majority of stars with $i' > 23.5$ in Region III$_g$ 
are evolving on the RGB, while those with $i' < 23.5$ are evolving on the AGB.
The M33 halo field contains a modest excess population of sources with respect 
to the control field when $i' > 21$, which is due to the RGB.}

\figcaption
[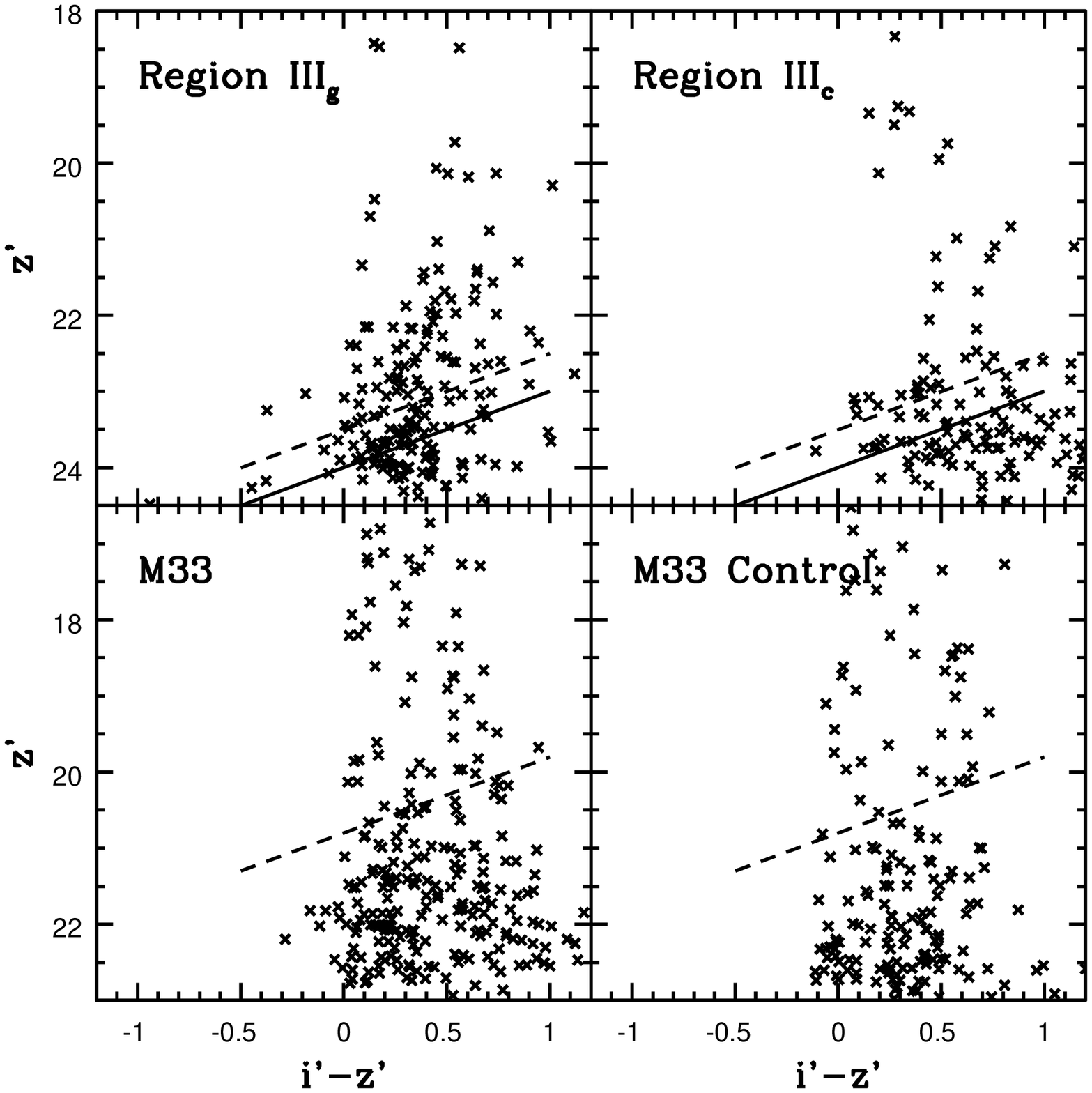]
{The $(z', i'-z')$ CMDs of the galaxy and control fields for NGC 2403 
Region III and M33. The dashed line in the upper panels 
marks $i' = 23.5$, which is the approximate brightness of the RGB-tip in NGC 
2403 with only foreground extinction, while the solid line marks $i' = 24$, 
which is the RGB-tip with an assumed internal extinction of A$_{g'} = 1$ mag. 
The overall morphologies of the $(z', i'-z')$ CMDs of NGC 2403 Regions I$_g$, 
II$_g$, and III$_g$ are very similar. The dashed line in the lower panels 
marks $i' = 20.8$, which is the approximate brightness of the RGB-tip 
in M33. The M33 halo field contains an excess population of sources with 
respect to the control field when $z' < 20$, and the majority of these are 
RGB stars. The M33 field also contains a modest excess population of sources 
above the RGB-tip, which are stars evolving on the AGB.}

\figcaption
[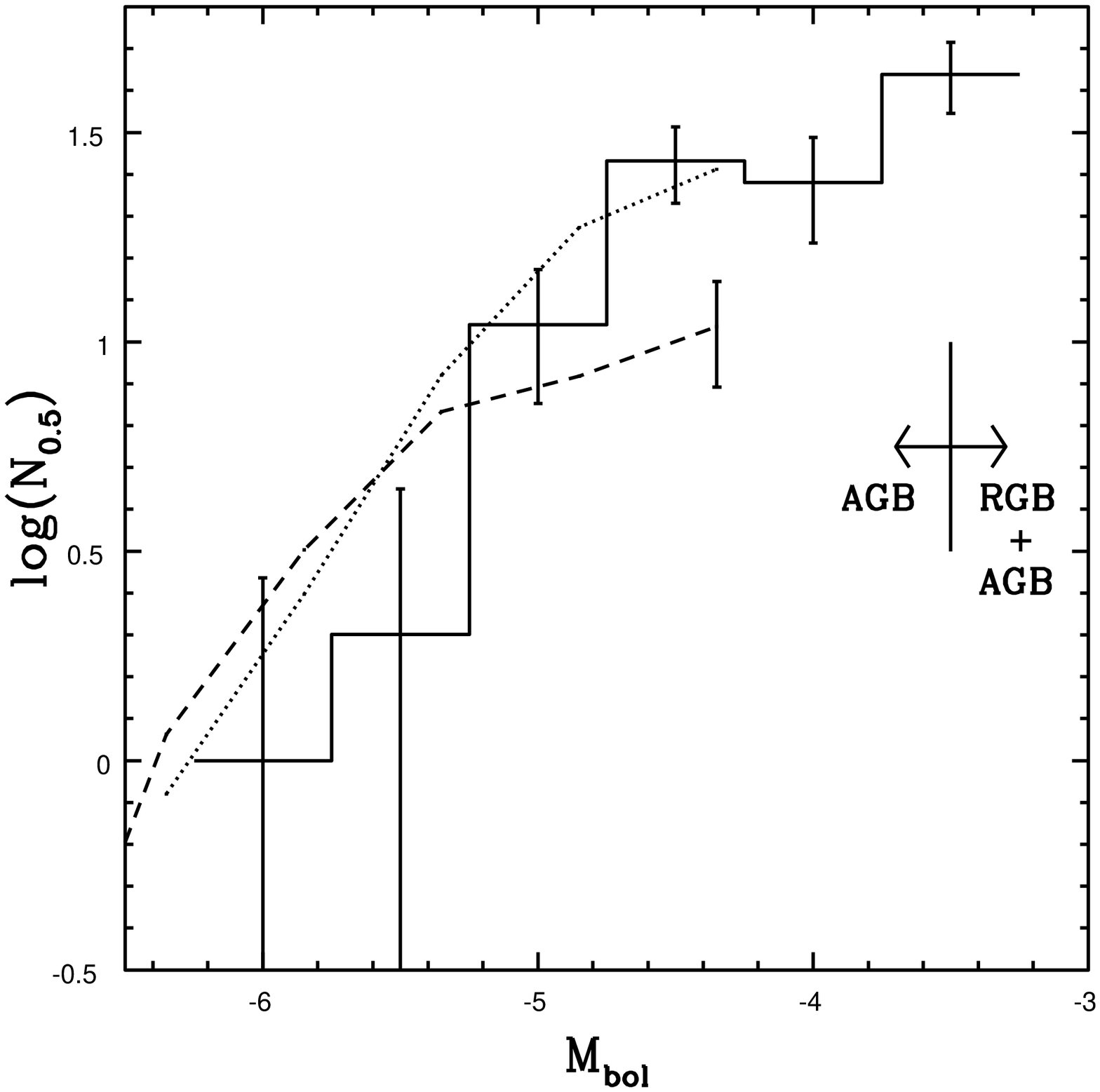]
{The LF of AGB stars in Region I$_g$ (solid line), 
corrected for foreground and background contamination, compared with the AGB 
LFs of Hudon et al. (1987) Fields 1 (dashed line) and 2 $+$ 3 (dotted line). 
The approximate magnitude of the RGB-tip is indicated with the vertical line; 
sources fainter than the RGB-tip are a combination of RGB and AGB stars, while 
sources brighter than the RGB-tip are on the AGB. 
The Hudon et al. (1989) LFs have been shifted along the vertical axis to match 
the surface brightness in Region I$_g$ using the 
measurements listed in Table 5 and also to correct for differences in 
areal coverage. The Hudon et al. LFs have also been shifted by 0.1 mag along 
the horizontal axis to account for the different distance modulus adopted for 
the current work. Note that the 3 LFs agree within the estimated 
errors when M$_{bol} < -5$, indicating that the space density of luminous AGB 
stars, when normalised to the total stellar content in each field, does not 
change markedly with radius in NGC 2403. The Hudon et al. Field 1 LF falls 
well below that of the Region I$_g$ and Hudon et al. Field $2+3$ LFs at 
M$_{bol} = -4.5$, although this may be due to incompleteness (see text). 
The errorbar at the faint end of the Field 1 LF shows the uncertainties 
listed in Table V of Hudon et al. (1989).}

\end{document}